\documentstyle[12pt,epsf]{article}

\begin{document}

\newcommand{\sst}[1]{{\scriptscriptstyle #1}}
\newcommand{\beq}{\begin{equation}}
\newcommand{\eeq}{\end{equation}}
\newcommand{\beqa}{\begin{eqnarray}}
\newcommand{\eeqa}{\end{eqnarray}}
\newcommand{\dida}[1]{/ \!\!\! #1}
\renewcommand{\Im}{\mbox{\sl{Im}}}
\renewcommand{\Re}{\mbox{\sl{Re}}}
\def\simge{\hspace*{0.2em}\raisebox{0.5ex}{$>$}
     \hspace{-0.8em}\raisebox{-0.3em}{$\sim$}\hspace*{0.2em}}
\def\simle{\hspace*{0.2em}\raisebox{0.5ex}{$<$}
     \hspace{-0.8em}\raisebox{-0.3em}{$\sim$}\hspace*{0.2em}}
\def\dn{{d_n}}
\def\de{{d_e}}
\def\datom{{d_\sst{A}}}
\def\grhobar{{{\bar g}_\rho}}
\def\gpibar{{{\bar g}_\pi^{(I) \prime}}}
\def\gpibarz{{{\bar g}_\pi^{(0) \prime}}}
\def\gpibaro{{{\bar g}_\pi^{(1) \prime}}}
\def\gpibart{{{\bar g}_\pi^{(2) \prime}}}
\def\mn{{m_\sst{N}}}
\def\mx{{M_X}}
\def\mrho{{m_\rho}}
\def\qpv{{Q_\sst{W}}}
\def\lamtv{{\Lambda_\sst{TVPC}}}
\def\lamtvs{{\Lambda_\sst{TVPC}^2}}
\def\lamtvc{{\Lambda_\sst{TVPC}^3}}

%     \hspace{-0.8em}\raisebox{-0.3em}{$\sim$}\hspace*{0.2em}}
\def\bra#1{{\langle#1\vert}}
\def\ket#1{{\vert#1\rangle}}
\def\coeff#1#2{{\scriptstyle{#1\over #2}}}
\def\undertext#1{{$\underline{\hbox{#1}}$}}
\def\hcal#1{{\hbox{\cal #1}}}
\def\sst#1{{\scriptscriptstyle #1}}
\def\eexp#1{{\hbox{e}^{#1}}}
\def\rbra#1{{\langle #1 \vert\!\vert}}
\def\rket#1{{\vert\!\vert #1\rangle}}

\def\lsim{{ <\atop\sim}}
\def\gsim{{ >\atop\sim}}
\def\nubar{{\bar\nu}}
\def\psibar{{\bar\psi}}
\def\Gmu{{G_\mu}}
\def\alr{{A_\sst{LR}}}
\def\wpv{{W^\sst{PV}}}
\def\evec{{\vec e}}
\def\notq{{\not\! q}}
\def\notl{{\not\! \ell}}
\def\notk{{\not\! k}}
\def\notp{{\not\! p}}
\def\notpp{{\not\! p'}}
\def\notder{{\not\! \partial}}
\def\notcder{{\not\!\! D}}
\def\notA{{\not\!\! A}}
\def\notv{{\not\!\! v}}
\def\Jem{{J_\mu^{em}}}
\def\Jana{{J_{\mu 5}^{anapole}}}
\def\nue{{\nu_e}}
\def\mn{{m_\sst{N}}}
\def\mns{{m^2_\sst{N}}}
\def\me{{m_e}}
\def\mes{{m^2_e}}
\def\mq{{m_q}}
\def\mqs{{m_q^2}}
\def\mw{{M_\sst{W}}}
\def\mz{{M_\sst{Z}}}
\def\mzs{{M^2_\sst{Z}}}
\def\ubar{{\bar u}}
\def\dbar{{\bar d}}
\def\sbar{{\bar s}}
\def\qbar{{\bar q}}
\def\sstw{{\sin^2\theta_\sst{W}}}
\def\gv{{g_\sst{V}}}
\def\ga{{g_\sst{A}}}
\def\pv{{\vec p}}
\def\pvs{{{\vec p}^{\>2}}}
\def\ppv{{{\vec p}^{\>\prime}}}
\def\ppvs{{{\vec p}^{\>\prime\>2}}}
\def\qv{{\vec q}}
\def\qvs{{{\vec q}^{\>2}}}
\def\xv{{\vec x}}
\def\xpv{{{\vec x}^{\>\prime}}}
\def\yv{{\vec y}}
\def\tauv{{\vec\tau}}
\def\sigv{{\vec\sigma}}

\def\sst#1{{\scriptscriptstyle #1}}
\def\gpnn{{g_{\sst{NN}\pi}}}
\def\grnn{{g_{\sst{NN}\rho}}}
\def\gnnm{{g_\sst{NNM}}}
\def\hnnm{{h_\sst{NNM}}}
\def\xivz{{\xi_\sst{V}^{(0)}}}
\def\xivt{{\xi_\sst{V}^{(3)}}}
\def\xive{{\xi_\sst{V}^{(8)}}}
\def\xiaz{{\xi_\sst{A}^{(0)}}}
\def\xiat{{\xi_\sst{A}^{(3)}}}
\def\xiae{{\xi_\sst{A}^{(8)}}}
\def\xivtez{{\xi_\sst{V}^{T=0}}}
\def\xivteo{{\xi_\sst{V}^{T=1}}}
\def\xiatez{{\xi_\sst{A}^{T=0}}}
\def\xiateo{{\xi_\sst{A}^{T=1}}}
\def\xiva{{\xi_\sst{V,A}}}
\def\rvz{{R_\sst{V}^{(0)}}}
\def\rvt{{R_\sst{V}^{(3)}}}
\def\rve{{R_\sst{V}^{(8)}}}
\def\raz{{R_\sst{A}^{(0)}}}
\def\rat{{R_\sst{A}^{(3)}}}
\def\rae{{R_\sst{A}^{(8)}}}
\def\rvtez{{R_\sst{V}^{T=0}}}
\def\rvteo{{R_\sst{V}^{T=1}}}
\def\ratez{{R_\sst{A}^{T=0}}}
\def\rateo{{R_\sst{A}^{T=1}}}
\def\mro{{m_\rho}}
\def\mks{{m_\sst{K}^2}}
\def\mpi{{m_\pi}}
\def\mpis{{m_\pi^2}}
\def\mom{{m_\omega}}
\def\mphi{{m_\phi}}
\def\Qhat{{\hat Q}}
\def\FOS{{F_1^{(s)}}}
\def\FTS{{F_2^{(s)}}}
\def\GAS{{G_\sst{A}^{(s)}}}
\def\GES{{G_\sst{E}^{(s)}}}
\def\GMS{{G_\sst{M}^{(s)}}}
\def\GATEZ{{G_\sst{A}^{\sst{T}=0}}}
\def\GATEO{{G_\sst{A}^{\sst{T}=1}}}
\def\mdax{{M_\sst{A}}}
\def\mustr{{\mu_s}}
\def\rsstr{{r^2_s}}
\def\rhostr{{\rho_s}}
\def\GEG{{G_\sst{E}^\gamma}}
\def\GEZ{{G_\sst{E}^\sst{Z}}}
\def\GMG{{G_\sst{M}^\gamma}}
\def\GMZ{{G_\sst{M}^\sst{Z}}}
\def\GEn{{G_\sst{E}^n}}
\def\GEp{{G_\sst{E}^p}}
\def\GMn{{G_\sst{M}^n}}
\def\GMp{{G_\sst{M}^p}}
\def\GAp{{G_\sst{A}^p}}
\def\GAn{{G_\sst{A}^n}}
\def\GA{{G_\sst{A}}}
\def\GETEZ{{G_\sst{E}^{\sst{T}=0}}}
\def\GETEO{{G_\sst{E}^{\sst{T}=1}}}
\def\GMTEZ{{G_\sst{M}^{\sst{T}=0}}}
\def\GMTEO{{G_\sst{M}^{\sst{T}=1}}}
\def\lamd{{\lambda_\sst{D}^\sst{V}}}
\def\lamn{{\lambda_n}}
\def\lams{{\lambda_\sst{E}^{(s)}}}
\def\bvz{{\beta_\sst{V}^0}}
\def\bvo{{\beta_\sst{V}^1}}
\def\Gdip{{G_\sst{D}^\sst{V}}}
\def\GdipA{{G_\sst{D}^\sst{A}}}
\def\fks{{F_\sst{K}^{(s)}}}
\def\FIS{{F_i^{(s)}}}
\def\fpi{{F_\pi}}
\def\fk{{F_\sst{K}}}
\def\RAp{{R_\sst{A}^p}}
\def\RAn{{R_\sst{A}^n}}
\def\RVp{{R_\sst{V}^p}}
\def\RVn{{R_\sst{V}^n}}
\def\rva{{R_\sst{V,A}}}
\def\xbb{{x_B}}
\def\mlq{{M_\sst{LQ}}}
\def\mlqs{{M_\sst{LQ}^2}}
\def\lscal{{\lambda_\sst{S}}}
\def\lvect{{\lambda_\sst{V}}}
\def\PR#1{{{\em   Phys. Rev.} {\bf #1} }}
\def\PRC#1{{{\em   Phys. Rev.} {\bf C#1} }}
\def\PRD#1{{{\em   Phys. Rev.} {\bf D#1} }}
\def\PRL#1{{{\em   Phys. Rev. Lett.} {\bf #1} }}
\def\NPA#1{{{\em   Nucl. Phys.} {\bf A#1} }}
\def\NPB#1{{{\em   Nucl. Phys.} {\bf B#1} }}
\def\AoP#1{{{\em   Ann. of Phys.} {\bf #1} }}
\def\PRp#1{{{\em   Phys. Reports} {\bf #1} }}
\def\PLB#1{{{\em   Phys. Lett.} {\bf B#1} }}
\def\ZPA#1{{{\em   Z. f\"ur Phys.} {\bf A#1} }}
\def\ZPC#1{{{\em   Z. f\"ur Phys.} {\bf C#1} }}
\def\etal{{{\em   et al.}}}
\def\delalr{{{delta\alr\over\alr}}}
\def\pbar{{\bar{p}}}
\def\lamchi{{\Lambda_\chi}}
\def\qw0{{Q_\sst{W}^0}}
\def\qwp{{Q_\sst{W}^P}}
\def\qwn{{Q_\sst{W}^N}}
\def\qwe{{Q_\sst{W}^e}}
\def\qem{{Q_\sst{EM}}}
\def\gae{{g_\sst{A}^e}}
\def\gve{{g_\sst{V}^e}}
\def\gvf{{g_\sst{V}^f}}
\def\gaf{{g_\sst{A}^f}}
\def\gvu{{g_\sst{V}^u}}
\def\gau{{g_\sst{A}^u}}
\def\gvd{{g_\sst{V}^d}}
\def\gad{{g_\sst{A}^d}}
\def\gvftil{{\tilde g_\sst{V}^f}}
\def\gaftil{{\tilde g_\sst{A}^f}}
\def\gvetil{{\tilde g_\sst{V}^e}}
\def\gaetil{{\tilde g_\sst{A}^e}}
\def\gvqtil{{\tilde g_\sst{V}^e}}
\def\gaqtil{{\tilde g_\sst{A}^e}}
\def\gvutil{{\tilde g_\sst{V}^e}}
\def\gautil{{\tilde g_\sst{A}^e}}
\def\gvdtil{{\tilde g_\sst{V}^e}}
\def\gadtil{{\tilde g_\sst{A}^e}}
\def\hvf{{h_\sst{V}^f}}
\def\hvu{{h_\sst{V}^u}}
\def\hvd{{h_\sst{V}^d}}
\def\hve{{h_\sst{V}^e}}
\def\hvq{{h_\sst{V}^q}}
\def\delp{{\delta_P}}
\def\delzp{{\delta_{00}}}
\def\deld{{\delta_\Delta}}
\def\dele{{\delta_e}}
\def\lnew{{{\cal L}_\sst{NEW}}}
\def\osffp{{{\cal O}_{7a}^{ff'}}}
\def\oszg{{{\cal O}_{7c}^{Z\gamma}}}
\def\osgg{{{\cal O}_{7b}^{g\gamma}}}

\begin{titlepage}

\hfill{INT \#DOE/ER/40561-41-INT98}

\vspace{1.0cm}

\begin{center}

{\large{\bf Constraints on T-Odd, P-Even Interactions
 from Electric Dipole Moments, Revisited}}

\vspace{1.2cm}

A. Kurylov$^1$,
G. C. McLaughlin$^{2,4}$, and M.J.  Ramsey-Musolf$^{1,3}$

\vspace{0.8cm}

$^{1}$ Department of Physics, University of Connecticut, Storrs, CT 06269 USA\\
$^{2}$ Department of Physics and Astronomy, State University of New York,
Stony Brook,
NY 11794 USA \\
$^{3}$ Theory Group, Thomas Jefferson National Accelerator Facility,
Newport News, VA
  23606 USA\\
$^{4}$ Previous address: TRIUMF, 4004 Wesbrook Mall, Vancouver, B.C. Canada
V6T2A3\\

\date{Preliminary draft}

\end{center}
\begin{abstract} We construct the relationship between nonrenormalizable,
effective, time-reversal
violating (TV) parity-conserving (PC) interactions of quarks and gauge
bosons and various
low-energy TVPC and TV parity-violating (PV) observables. Using effective
field theory
methods, we delineate the scenarious under which experimental limits on
permanent electric
dipole moments (EDM's) of the electron, neutron, and neutral atoms as well
as limits on
TVPC observables provide the most stringent bounds on new TVPC
interactions. Under scenarios in
which parity invariance is restored at short distances, the one-loop EDM of
elementary fermions
generate the most severe constraints. The limits derived from the atomic
EDM of $^{199}$Hg are
considerably weaker. When parity symmetry remains broken at short
distances, direct TVPC
search limits provide the least ambiguous bounds. The direct limits follow
from TVPC interactions
between two quarks.

\end{abstract}

\end{titlepage}

\section{Introduction}

The search for physics beyond the Standard Model (SM) is a topic of
considerable interest in
high-energy particle physics. Concurrently, efforts are also underway to
uncover signatures of new
physics at low- and medium-energies using atomic and nuclear processes. In
this respect, there
exist tantalizing hints of new physics in results of neutron and
superallowed nuclear
$\beta$-decays, which imply a value for $|V_{ud}|$ differing from the SM
unitarity requirement by
two or more $\sigma$ \cite{PDG00}. Similarly, the weak charge of the cesium
atom,
$\qpv$, measured in
atomic parity-violation (APV) by the Boulder group, has been found to
differ from the SM
prediction by $2.5\sigma$ \cite{Ben99} (See, however, Ref.
\cite{Der00}). If conventional many-body atomic and nuclear
effects can be ruled out
as the source of these deviations, the $\beta$-decay and APV results imply
the existence of new
physics at the 1-10 TeV scale\cite{MRM99b,MRM00a}. This possibility has
motivated
a variety of additional
atomic and nuclear new physics searches, including new measurements of the
neutron $\beta$-decay
parameters, APV observables along a chain of isotopes, and parity-violating
(PV) electron-electron and electron-proton scattering.

One possible manifestation of new physics not probed by the aforementioned
experiments would be the
existence of new low-energy interactions involving a single generation of
fermions
which violate time-reversal invariance (T) but conserve
parity invariance (P). Such interactions are allowed in the SM when quarks
of different generations participate.
Recently, the first non-zero result for a $\Delta S=1$
T-violating, P-conserving (TVPC)
observable has been reported by the CPLEAR Collaboration, which measured
the $K^0-{\bar K}^0$
decay asymmetry \cite{Ang98}.
The results are consistent with the value expected from the measured
CP-violating parameter
$\epsilon$ and the CPT theorem. No new physics is required to explain this
result. In the $\Delta
S=0$ sector, a variety of direct searches for TVPC effects have been
carried out. These efforts
include studies of detailed balance in nuclear reactions\cite{Bla83},
$\gamma$-ray correlations in
nuclear $\gamma$-decay\cite{Boe95}, five-fold correlations (FC) in the
scattering of
epithermal neutrons from
aligned nuclear targets\cite{Huf97,Koe91}, charge symmetry breaking (CSB) in
$np$ scattering \cite{Sim97,COSY98,Van99},
and the ${\hat J}\cdot ({\hat
p_e}\times{\hat p_\nu})$ in neutron
$\beta$-decay\cite{Lis00}. Thus far, all studies have yielded null results. The
limits from the purely
hadronic reactions imply $\alpha_T\simle\ {\hbox{few}}\times 10^{-3}$,
where $\alpha_T$ gives the
ratio of TVPC nuclear matrix elements to those of the residual strong
interaction.

Limits on $\Delta S=0$ TVPC interactions involving light quarks may also be
derived indirectly
from results for atomic,
neutron, and electron electric dipole moments (EDM's). As observed in Ref.
\cite{Khr91}, the
presence of both a new TVPC interaction and a conventional PV interaction
({\em e.g.}, in the
Standard Model) could conspire to generate a non-zero EDM, whose
interaction with an external
field violates both P and T. To the extent that PV radiative
corrections to possible new TVPC
interactions can be calculated, one can derive limits on new TVPC
interactions from EDM results.
Attempts to do so were first reported in Ref. \cite{Khr91}. The calculation
involved two external
elementary fermions ({\em e.g.}, two valence quarks in the neutron) and a
one-loop $Z$-boson
radiative correction to dimension seven, four-fermion, TVPC operators.
Two-loop effects,
involving a single external
fermion, for the electron EDM, $d_e$, and neutron EDM, $\dn$, were later
studied in Ref.
\cite{Con92}.  Na\"\i vely, one might expect the most stringent bounds
on new TVPC interactions to
be derived at one-loop
order from the experimental limit on the atomic electric dipole moment of
mercury,
$\datom(^{199}{\hbox{Hg}})$, since the latter is nearly two orders of
magnitude more severe than
the bound on $\dn$. It was argued in Ref. \cite{Con92}, however, that the
two-loop effects in
$d_e$ and $\dn$ generate considerably more stringent bounds than do the
results of $\datom$.
Subsequently, the authors of Ref. \cite{Eng96} recast the analysis of Refs.
\cite{Khr91,Con92}
into the framework of low-energy effective field theory (EFT). It was
argued in Ref. \cite{Eng96} that the results of Ref. \cite{Con92} imply
bounds on new
TVPC interactions in
excess of those presently achievable with direct TVPC searches by several
orders of magnitude.
These conclusions have had a discouraging effect on further direct TVPC
searches.

Recently, it was argued that the conclusions of Ref. \cite{Con92} are
inconsistent with the separation of scales underlying EFT
\cite{MRM99a}. In brief, the argument is as follows\cite{MRM00b}.
Let $\lamtv$ denote the mass scale
below which use of an EFT involving nonrenormalizable TVPC operators makes
sense.
One may expand the EDM of an elementary particle, neutron, or atom as
\begin{equation}
\label{eq:edmexpand}
d = \beta_5 C_5{1\over\lamtv}+\beta_6C_6{M\over\lamtvs}+\beta_7C_7{M^2\over
\lamtvc}+\cdots\ \ \ ,
\end{equation}
where the $C_d$ denote the set of {\em a priori} uknown coefficients of
dimension $d$
nonrenormalizable operators in the effective Lagrangian, the $\beta_d$ are
calculable
quantities arising from loops or many-body matrix elements, and $M<\lamtv$
is a mass
scale associated with the appropriate dynamical degree of freedom in the
EFT. The $C_d$
parameterize one's ignorance about the short-distance ($p\simge\lamtv$)
dynamics of the
new time-reversal violating physics. The first contributions from new TVPC
interactions
appear in the $C_7$.

\noindent One may now consider Eq. (\ref{eq:edmexpand}) under two scenarios:

\medskip
\noindent {\bf Scenario (A)} Parity symmetry is restored at some scale
$\mu\simle\lamtv$.
In this case, all of the coefficients $C_5$ and $C_6$ must vanish at
tree-level in the
EFT since parity invariance holds at short distances. Consequently, the
first contributions
to the EDM arise from loops involving the TVPC $C_7$ operators. Since
$M/\lamtv<1$, these
contributions presumably dominate the remaining terms in the series. Hence,
one may use
experimental EDM limits to constrain $C_7/\lamtvc$. As shown below, the
limits obtained
from EDM's under this scenario vastly exceed those obtainable from direct
searches.

\medskip
\noindent {\bf Scenario (B)} Parity symmetry is restored at
$\mu\simge\lamtv$. In this
case, the $C_5$ and $C_6$ do not, in general, vanish at tree-level in the
EFT since
both PV and TV interactions take place at short distance. There exists no
reason to assume they fail to conspire in generating the lowest dimension TVPV
effective interactions. Consequently, the TVPC interactions do not generate
the leading
contribution to the EDM as in Scenario (A). Without independent information
on the $C_{5,6}$
one cannot use the EDM as a direct handle on the TVPC $C_7$ terms. The
latter may be more or
less suppressed relative to the lower dimension contributions depending on
the size of
$M/\lamtv$. Since one has no {\em a priori} information on $M/\lamtv$, one
can say very
little about the importance of TVPC contributions. For the sake of
argument, one might
assume $M/\lamtv << 1$ so that the first term in Eq. (\ref{eq:edmexpand})
dominates. In
this case, the low-energy effects of TVPC interactions would be negligible.
In the more
general situation, however, one would have to use direct TVPC searches to
constrain the
new TVPC interactions under this scenario.

The analysis of Refs. \cite{Con92,Khr91} implicitly assumes Scenario (A).
The EDM
calculations performed by these authors, however, do not display the proper
$1/\lamtvc$ scaling behavior which follows from EFT. The reasons for this
failure
are discussed in Ref. \cite{MRM99a} and summarized below. It was also
shown in Ref. \cite{MRM99a} that
there exist additional TVPC operators, not considered in Refs.
\cite{Khr91,Con92,Eng96},
which contribute to the elementary fermion EDM at one-loop order. Under
Scenario (A),
these one loop effects yield the most stringent constraints on the size of
TVPC effects.

In what follows, we extend the analysis of Ref. \cite{MRM99a} to include
many-quark TVPC contributions to the neutron EDM -- first studied in
Ref. \cite{Khr91} -- and to atomic EDM's. We concentrate on Scenario (A),
since under
Scenario (B) one cannot use EDM's to derive unambiguous information about
TVPC new physics.
In the case of the neutron EDM,
we complete the one-loop analysis of Ref. \cite{Khr91}, including additional diagrams required by electromagnetic gauge invariance. We
show that
the impact of these new diagrams is as large as the one-loop effects
considered
previously in Ref. \cite{Khr91}. We also compute tree-level contributions
arising
from dimension seven TVPC operators not considered in Ref. \cite{Khr91}.

In the case of atomic EDM's, we consider the situation in which they
arise from purely hadronic TVPC interactions in the nucleus.
Traditionally, the effects of non-leptonic T-violation in nuclear and
atomic processes have been analyzed using collective degrees of freedom
(mesons and baryons), rather than fundamental
quark-quark or quark-gluon interactions. The T-violating effects are
characterized by hadronic coupling constants which may
be related to the underlying quark and gluon T-violating interactions using
standard hadron structure techniques.
In this context, two hadronic effects are of interest: (a) the presence of
a purely TVPC meson-nucleon interaction and (b) the presence of a TVPV
meson-nucleon
interaction.

The leading \lq\lq long-range" TVPC effect arises from $\rho$-meson
exchange, where the TVPC $\rho NN$ vertex is characterized by a coupling
strength $\grhobar$ and an interaction \cite{Her95}
\begin{equation}
\label{eq:lrhonn}
{\cal L}_{\rho NN}^\sst{TVPC} = i\sqrt{2} \grhobar f_\rho {\kappa_V \over 2
m_n}
\bar{N} \sigma^{\mu \lambda}(\tau^- \partial_\lambda\rho^+_\mu - \tau^+
\partial_\lambda\rho_\mu^-) N
\ \ \ ,
\end{equation}
where $f_\rho=2.79$ and $\kappa_V=3.70$.
A time-reversal violating parity violating (TVPV) nuclear effect arises
when the second vertex in the exchange is parity-violating. Alternately,
a TVPV atomic moment can be generated by a TVPC nuclear $\rho$-exchange and
the PV exchange of a $Z$-boson between the nucleus and atomic
electrons (see Fig. \ref{fig:atomic}).

Similarly, the longest-range TVPV effects generally arise from
$\pi$-exchange. In this case, the relevant TVPV $\pi NN$ couplings are
$\gpibar$, where the superscript denotes the isospin channel and correspond
to the interactions
\begin{eqnarray}
\label{eq:lpinn}
{\cal L}_{\pi NN}^\sst{TVPV}&=&{\bar N}\biggl[ \gpibarz{\vec\tau}\cdot{\vec\pi}
+\gpibaro\pi^0\\
\nonumber
&&+\gpibart (3\tau_z\pi^0-{\vec\tau} \cdot{\vec \pi} )\biggr] N
\end{eqnarray}
Non-zero values of $\gpibar$ may arise from either new TVPC interactions plus
weak radiative corrections, or from more conventional TVPV interactions, such
as the $\theta$-term in the QCD Lagrangian. The latter have been considered
extensively elsewhere \cite{Her95}, and we concentrate on the former.

Upper bounds on $|\grhobar|$ have been derived from a variety of
T-violating, P-conserving
experiments, including  the studies of detailed balance in nuclear
reactions\cite{Bla83}, neutron
transmission through an aligned $^{165}$Ho target\cite{Huf97}, and
CSB terms in the $np$ scattering cross section\cite{Sim97}. In addition,
measurements of $\dn$ and
$\datom$ yield bounds on $\grhobar$, when  PV and  TVPC interactions
conspire to generate an EDM. Under Scenario (A), the EDM limits on
$\grhobar$ may
be stronger than those obtained from
TVPC nuclear processes\cite{horing}. Regarding the
$\gpibar$, experimental
EDM limits yield the most stringent bounds. As discussed in detail in Ref.
\cite{Her95},
$\dn$ and $\datom(^{199}{\hbox{Hg}})$ are sensitive to different linear
combinations
of the $\gpibar$. Roughly speaking, the EDM upper bounds on the $|\gpibar|$
are of the order of a few $\times 10^{-11}$.

In this paper, we relate $\grhobar$ and the $\gpibar$ to the underlying
TVPC quark and gluon
operators, and use limits on the hadronic couplings to infer limits on
these underlying
interactions. We then compare these limits as well as those obtained
from the one-loop many-quark contributions to $\dn$ with those obtained
from the EDM's of
elementary fermions as in
Ref. \cite{MRM99a}. We find:

\begin{enumerate}

\item Under Scenario (A), the one-loop elementary fermion EDM studied
in Ref. \cite{MRM99a} produces the
most stringent limits on new TVPC interactions when applied to the electron
and neutron.
Many-quark effects $d_n$ or $d_A$ are suppressed by at least five orders of
magnitude.

\item By making certain naturalness assumptions, one may use experimental
EDM limits
to derive lower bounds on $\lamtv$. Under Scenario (A), one infers from
$d_e$ (single-quark
$\dn$) limits that
$\lamtv\simge 260$ TeV ($\sim 110$ TeV) if the new TVPC physics is strong.
The corresponding bounds derived from many-quark effects in $\dn$
and $\datom(^{199}\mbox{Hg})$ are at roughly 1000 times weaker. In order
for $\datom(^{199}
\mbox{Hg})$ to provide competitive limits, the precision of the atomic EDM
experiments
would need to improve by roughly nine orders of magnitude\footnote{However,
at some
point, improved precision in $\datom$ would improve tighten the bounds on
$\dn$.}.

\item Under Scenario (A), one expects $\alpha_T\simle 10^{-15}$.

\item For Scenario (B), experimental limits on $\grhobar$ derived from FC
in neutron
transmission and CSB in $np$ scattering produce the strongest limits on
TVPC new physics.
In terms of mass scales, these limits give $\lamtv\simge 1$ GeV for new
strong TVPC physics.
An improvement of six orders of magnitude in experimental precision would
bring this
lower bound up to the weak scale.

\end{enumerate}

The analysis leading to these conclusions is presented in the remainder of
the paper as follows.  In section 2, we review the framework of EFT
for new low-energy TVPC and TVPV interactions.
In section 3 we illustrate the application of this framework by
considering the EDM of an
elementary fermion, as discussed in Ref. \cite{MRM99a}. In section 4, we
discuss
the renormalization of
effective TVPV four-fermion operators arising from PV radiative corrections
to effective TVPC
interactions. Here, we take particular care to implement electromagnetic
gauge invariance. The
latter implies the existence of additional contributions to the EDM's of
composite systems not
considered in Ref.
\cite{Khr91}. In section 5, we  relate the effective TVPC and
TVPV operators to
$\grhobar$ and the $\gpibar$, respectively, using the quark model,
factorization, and current
algebra. We also compute new many-quark contributions to $\dn$ generated
by new operators -- including those required by gauge invariance -- not
considered in Ref. \cite{Khr91}.
In section 6, we compare the
implications of the many-quark contributions for the scale of new TVPC
interactions with
those obtained from the study of Ref. \cite{MRM99a}. We also consider the
limits on $\grhobar$ and
the $\gpibar$ obtained from atomic EDM limits and direct searches
and the corresponding
implications for new TVPC interactions under the two scenarios outlined
above.
Section 6 summarizes our conclusions.

\section{Effective field theory and new TVPC interactions}

Herczeg {\em et al} have shown that TVPC interactions between
quarks cannnot arise
from tree-level boson exchange
in renormalizable gauge theories \cite{Her92}. Such
interactions could, however,
be generated by higher-order or non-perturbative effects.
Whatever new physics
produces P-conserving T-violation among light quarks and gluons must be
characterized by some
heavy mass scale, $\lamtv$. Given that the underlying renormalizable gauge
theory for P-conserving
T-violation is not known, it is natural to consider the low-energy
consequences of such a theory
in the context of an EFT valid below the scale $\lamtv$.
Letting $\lnew$ denote
the effective, low-energy Lagrangian for new physics, we follow Ref.
\cite{Eng96} and expand
in inverse powers of $\lamtv$:
\begin{equation}
\label{eq:lnew}
\lnew = {\cal L}_4+{1\over\lamtv}{\cal L}_5 +{1\over\lamtvs}{\cal L}_6
+{1\over\lamtvc}{\cal L}_7 +\cdots\ \ \ ,
\end{equation}
where the subscripts denote the dimension of the operators appearing in
each term and where
\begin{equation}
\label{eq:lnewexpand}
{\cal L}_d = \sum_k C_d^k {\cal O}_d^k
\end{equation}
with the sum running over $k$ complete set of dimension $d$ operators
$\{{\cal O}_d^k\}$. The lowest dimension T-violating operators appear in
${\cal L}_5$. These operators are TVPV only. Of particular interest
here are the electric dipole fermion-gauge boson interactions:
\begin{eqnarray}
{\cal O}_{5}^{f\gamma}&=& -{i\over
2}{\bar\psi}_f\sigma_{\mu\nu}\gamma_5\psi_f\
F^{\mu\nu}\\
{\cal O}_{5}^{f Z}&=& -{i\over 2}{\bar\psi}_f\sigma_{\mu\nu}\gamma_5\psi_f\
Z^{\mu\nu}\\
{\cal O}_{5}^{f g}&=& -{i\over
2}{\bar\psi}_f\sigma_{\mu\nu}\gamma_5\lambda^a\psi_f\
G^{a\ \mu\nu}\ \ \ ,
\end{eqnarray}
where $F^{\mu\nu}$, $Z^{\mu\nu}$, and $G^{a\ \mu\nu}$ denote the photon,
$Z$-boson, and gluon
field strength tensors, respectively, and the $\lambda^a$ are the
Gell-Mann matrices.

The term ${\cal L}_6$ contains the lowest-order TVPV four-fermion
operators, which include
\begin{eqnarray}
\label{eq:d6tvpv}
{\cal O}_{6a}^{ff'}&=&i{\bar\psi}_f\psi_f
{\bar\psi}_{f'}\gamma_5\psi_{f'}\\
{\cal O}_{6b}^{ff'}&=&i{\bar\psi}_f\lambda^a\psi_f
     {\bar\psi}_{f'}\lambda^a\gamma_5\psi_{f'}\\
{\cal O}_{6c}^{ff'}&=&i{\bar\psi}_f\sigma_{\mu\nu}\psi_f
{\bar\psi}_{f'}
	   \sigma^{\mu\nu}\gamma_5\psi_{f'}\\
{\cal O}_{6d}^{ff'}&=&i{\bar\psi}_f\lambda^a\sigma_{\mu\nu}\psi_f
     {\bar\psi}_{f'}\lambda^a\sigma^{\mu\nu}\gamma_5\psi_{f'}
\end{eqnarray}
and so forth.

The lowest-dimension TVPC interactions arise in ${\cal L}_7$. Here, we
consider the
following three:
\begin{eqnarray}
\label{intro:l7a}
{\cal O}_{7a}^{ff'}&=&i \bar{\psi}_f \gamma_5 \sigma_{\mu \nu}
(\overleftarrow{D}_\nu + \overrightarrow{D}_\nu)
  \psi_f \bar{\psi}_{f'}\gamma^\mu \gamma_5 \psi_{f'} + \ {\hbox{h.c.}}\\
\label{intro:l7b}
  {\cal O}_{7b}^{g\gamma} &=& \bar{\psi}_f
\sigma_{\mu\nu}\lambda_a \psi_f
G^{\mu\alpha}_a F^{\nu}_{\alpha}\\
\label{intro:l7c}
 {\cal O}_{7c}^{Z\gamma} &=&  \bar{\psi}_f \sigma_{\mu\nu}
\psi_f
Z^{\mu\alpha} F^{\nu}_{\alpha}\ \ \ .
\end{eqnarray}
The operator $\osffp$ was first considered in Refs. \cite{Khr91,Con92},
while $\osgg$ was introduced in Ref. \cite{Eng96}. The interaction $\oszg$
was subsequently considered in Ref. \cite{MRM99a}.

The $d=7$ Lagrangian also contains several TVPV operators. Among those
relevant to us are
\begin{eqnarray}
\label{eq:d7tvpv}
{\cal O}_{7d}^{ff'}&=&{\bar{\psi_f}} \gamma^\mu
(\overleftarrow{D}_\mu -
	   \overrightarrow{D}_\mu) \psi_f \bar{\psi}_{f'} \gamma_5 \psi_{f'}\\
{\cal O}_{7e}^{ff'}&=&\bar{\psi_f} \psi_f \bar{\psi}_{f'}
\gamma^\mu \gamma_5
    (\overleftarrow{D}_\mu + \overrightarrow{D}_\mu) \psi_{f'}\\
{\cal O}_{7f}^{ff'}&=&i \bar{\psi_f} \gamma^\nu \psi_f
\bar{\psi}_{f'} \sigma_{\mu
    \nu} \gamma_5  (\overleftarrow{D}_\mu + \overrightarrow{D}_\mu) \psi_{f'}\\
{\cal O}_{7g}^{ff'}&=&i \bar{\psi_f} \gamma^\nu \gamma_5 \psi_f
\bar{\psi}_{f'}
    \sigma_{\mu \nu}  (\overleftarrow{D}_\mu - \overrightarrow{D}_\mu)
\psi_{f'}\\
{\cal O}_{7h}^{ff'}&=&\bar{\psi_f} \gamma^\mu \psi_f
\bar{\psi}_{f'} \gamma_5
    (\overleftarrow{D}_\mu - \overrightarrow{D}_\mu)\psi_{f'}\\
{\cal O}_{7i}^{ff'}&=&i \bar{\psi_f} \gamma^\mu
(\overleftarrow{D}^\nu +
     \overrightarrow{D}^\nu) \psi_f \bar{\psi}_{f'} \sigma_{\mu \nu}
\gamma_5  \psi_{f'}\\
{\cal O}_{7j}^{ff'}&=&i \bar{\psi_f} \gamma^\nu \gamma_5
(\overleftarrow{D}_\mu -
     \overrightarrow{D}_\mu) \psi_f \bar{\psi}_{f'} \sigma_{\mu \nu}
\psi_{f'}\ \ \ .
\end{eqnarray}
The specific forms of other $d\geq 5$ TVPV operators and $d\geq 7$ TVPC
operators are not relevant to the present discussion, so we do not list them
explicitly.

A key ingredient underlying the expansion of Eq. (\ref{eq:lnew})
is a separation of scales and an associated power-counting scheme.
Specifically,
the contribution to a T-violating observable from any physics associated
with scales
$\mu\simge\lamtv$ is contained in the operator
coefficients, $C_d$. These short-distance contributions are not calculable
since the underlying theory ({\em e.g.}, renormalizable gauge theory)
responsible for
TVPC effects is not
known\footnote{If this underlying theory were known, one would have no need
to employ the
expansion in Eq. (\ref{eq:lnew}) in the first place}. Consequently, the
$C_d$ can only be
determined from experiment. Contributions from physics having $\mu <\lamtv$
live in loops
and many-body ({\em e.g.}, hadronic) matrix elements containing the
non-renormalizable operators
${\cal O}_d$ and physical degrees of freedom having masses and momenta less
than the scale
$\lamtv$. Only these \lq\lq long-distance" contributions can be computed
using the EFT.

As a consequence of this scale separation, one obtains a systematic
power-counting scheme by
which to organize contributions to any low-energy T-violating observable.
If $d$ is the lowest
dimension of an effective operator which contributes to such an observable
${\cal A}^\sst{T-ODD}$, then
\begin{equation}
{\cal A}^\sst{T-ODD}\sim C_d\ \left({p\over\lamtv}\right)^{d-4}+\cdots \ \ \ ,
\end{equation}
where $p$ denotes a mass or momentum smaller than $\lamtv$. To the extent
that $p<<\lamtv$,
contributions from higher-dimension operators will be suppressed relative
to those from
${\cal O}_d$ by powers of $(p/\lamtv)$. In general, one may thus truncate
the expansion of ${\cal A}^\sst{T-ODD}$ at the
first or first few orders in $(p/\lamtv)$. In our
analysis of the EDM,
we find  $p$ is one of the following: elementary
fermion mass, $m_f$;
weak gauge boson mass, $\mz$; QCD scale, $\Lambda_\sst{QCD}$; inverse
hadron size,
$1/r_\sst{HAD}$; and long wavelength photon momentum, $q$.

Any renormalization of the ${\cal O}_d$ by loops involving any one of the
${\cal O}_d$ and,
{\em e.g.}, degrees of freedom appearing in ${\cal L}_4$ must be carried
out in a manner which
preserves the EFT scale separation. In particular, only intermediate states
with momenta and
energies  below $\lamtv$ must contribute to loop integrals. Higher-momentum
($p\simge\lamtv$) states
have been effectively integrated out in arriving at the expansion in Eq.
(\ref{eq:lnew}).
Consequently, we regulate all loop integrals using dimensional
regularization (DR), which
preserves the separation of scales. Since the subtraction scale $\mu$
arising in DR only
appears logarithmically in any regulated amplitude and never as a power,
the use of DR does not
alter the EFT power counting described above.
We emphasize that the cut-off regulator used in Refs. \cite{Khr91,Con92}
does not preserve the
EFT scale separation. In those analyses, loop integrals were cut off at
momenta $p\sim\lamtv$.
Consequently, in Refs. \cite{Khr91,Con92}
the renormalization of the EDM ${\cal O}_{5a}$ due to loops
involving any $d>5$
operator scales as $1/\lamtv$ and not as
$(p/\lamtv)^{d-5}\times (1/\lamtv)$ as implied by
EFT power counting. As we show below, this loss of power counting prevents
one from deriving
any information about the $d=7$ operators from experimental EDM limits. In
this respect, our
conclusions differ dramatically from those of Refs. \cite{Khr91,Con92}. We
illustrate this
point in the following section.

\bigskip

\section{Elementary fermion EDM}

The application of EFT for new TVPC interactions to the EDM of an
elementary fermion was
considered in Ref. \cite{MRM99a}. In what follows, we summarize the
arguments of that analysis, as
they illustrate the general principles of EFT to be used in the remainder
of this study. To that
end, we first observe that if, as in Scenario B, both new TVPC interactions
and PV
interactions ({\em e.g.}, in the
Standard Model) exist at momentum scales $p\gsim\lamtv$, then there
exists no reason to assume
that the coefficients of the TVPV effective operators in Eq.
(\ref{eq:lnew}) vanish at tree
level. Although we have no detailed knowledge of the dynamics of
short-distance TVPC and TCPV
interactions, nothing prevents their conspiring to generate non-vanishing
low-energy TVPV
interactions. In particular, the coefficients of the $d=5$ electric dipole
operators,
$C_5^{f\gamma}$ should be non-vanishing at tree-level unless some
fortuitous fine-tuning of the
short-distance TVPC and TCPV interactions occurs.

The situation here is analogous to the chiral expansion of the octet baryon
magnetic moments. In
the latter case, the leading order contribution occurs at tree level from
the $d=5$ magnetic
moment Lagrangian \cite{MRM97}:
\begin{eqnarray}
\label{eq:lmmoments}
{\cal L}_\sst{M.M} &=& {e\over
2\Lambda_\chi}\epsilon_{\mu\nu\alpha\beta}v^\alpha\big\{
b_{+}{\hbox{Tr}}\left({\bar B}_v S^\beta \{\lambda_3+\lambda_8/\sqrt{3},
B_v\}\right)\\
\nonumber
	& & +b_{-}{\hbox{Tr}}\left({\bar B}_v
[\lambda_3+\lambda_8/\sqrt{3}, B_v]\right)\ F^{\mu\nu}
\ \ \ ,
\end{eqnarray}
where the $B_v$ are the octet baryon fields for states of velocity
$v^\alpha$, $S^\alpha$ is
the spin operator, and $\Lambda_\chi$ is the scale of
chiral symmetry breaking.
The tree level relationship between a baryon magnetic
moment and the
low-energy constants $b_{\pm}$ is
\begin{equation}
\label{eq:mmoments}
\mu^a = \left({ m_\sst{B}\over\Lambda_\chi}\right)b^a\ \ \ ,
\end{equation}
where $m_\sst{B}$ is the mass of the baryon, \lq\lq $a$" denotes its SU(3)
indices, and $b^a$ is
the appropriate combination of the $b_{\pm}$. Since the baryon magnetic
moments are typically of
order unity, the tree-level relation implies that the $b_{\pm}$ are also of
order unity.
Alternatively, one may use Eq. (\ref{eq:mmoments}) to estimate the scale
$\Lambda_\chi$. If the
low-energy constants in Eq. (\ref{eq:lmmoments}) are of order unity, the
one must have
$\Lambda_\chi\sim m_\sst{B} \sim 1$ GeV.
Since the chiral symmetry of pionic interactions implies that
$\Lambda_\chi = 4\pi F_\pi\approx 1$ GeV,
the tree-level magnetic moment relation produces
a self-consistent value
for the scale of chiral symmetry breaking when the leading low-energy
constants are chosen to be
of order one.

In a similar way, one may use the tree-level relation between the EDM and
the coefficients of the
appropriate $d=5$ operators to estimate the scale $\lamtv$. This relation is
\begin{equation}
\label{eq:edmtree}
d_f = { C_5^{f\gamma}\over\lamtv} \ \ \ .
\end{equation}
We follow a standard convention for parameterizing the strength of new
physics interactions and
take $C_5^{f\gamma} = 4\pi\kappa^2 e$. Using the present limit for the EDM of
the electron
$|d_e| < 4\times 10^{-27}$ e-cm\cite{PDG00,Com94}, one obtains from Eq.
(\ref{eq:edmtree}) the
limit
$\lamtv > 10^{14}\ \kappa^2$ GeV. Thus, if the new TVPC physics is \lq\lq
strong"
($\kappa^2\sim 1$),  one
obtains a tremendously large value for the corresponding mass scale.

For both EFT's in Eqs. (\ref{eq:lnew}, \ref{eq:lmmoments}),
loop corrections
involving light, dynamical degrees of freedom modify the tree-level
relations in Eqs.
(\ref{eq:mmoments},\ref{eq:edmtree}). Single pion loop amplitudes, for
example, generate
corrections to isovector magnetic moments of ${\cal O}(\mpi/\Lambda_\chi)$
relative to the
tree-level contribution. The $\pi$ loops are quadratically divergent, yet
generate a finite
contribution to the $d=5$ magnetic moment operator. Power-counting implies
the appearance of one
additional mass factor in this finite contribution. When DR is used to
regulate the integral,
this mass factor becomes $\mpi$, resulting in the $\mpi/\Lambda_\chi$
suppression relative to
the tree level relation. This scaling behavior of the loop
contributions, which follows from the EFT
separation of scales,
provides for a chiral expansion in powers of $p/\Lambda_\chi$ (where
$p <\Lambda_\chi$) which
may be reasonably truncated at any order.
Similarly, if the EFT of Eq. (\ref{eq:lnew}) is well-behaved, one would
expect the
loop corrections to the relation in Eq.
(\ref{eq:edmtree}) to be
suppressed by powers of $p/\lamtv$, where $p < \lamtv$ is a mass scale
associated with the
dynamical degrees of freedom in the ${\cal L}_\sst{NEW}$.

For purposes of illustration, we assume that the only dynamical degrees of
freedom operative below $\lamtv$ are those appearing in the Standard
Model. Under Scenario A, additional degrees of freedom -- such
as the right-handed neutral gauge boson in left-right symmetric
theories -- would also generate loop contributions. These additional
degrees of freedom would be required in order for parity symmetry to
be restored for $\mu<\lamtv$. As noted below, however, the study
of only the standard model contributions yields conservative
upper bounds on the $d=7$ TVPC operators. In order to avoide introducing
model-dependence associated with parity restoration scenarios, we
restrict our attention to these SM effects. In this case,
the leading loop corrections
are generated by PV Standard Model radiative corrections to
the $d=7$ operators
${\cal O}_{7a-c}$. These corrections have been computed in Ref.
\cite{MRM99a} for ${\cal O}_{7a,c}$ using the diagrams in Figs.
\ref{fig:oneloop_2quark} and \ref{fig:twoloop}.
All of the loop amplitudes corresponding to Figs. \ref{fig:oneloop_2quark}
and \ref{fig:twoloop} are superficially
quadratically
divergent. The loops are regulated using DR and the poles removed by the
corresponding
counterterm in the renormalized $C_5^f$ in the $\overline{\hbox{MS}}$
subtraction
scheme. The closed
fermion loop subgraph in Fig. \ref{fig:twoloop}a  is nominally linearly
divergent and
corresponds to the
Adler-Bell-Jackiw anomaly diagram. In this case, the vector current
insertions arise from the
$\gamma$ and $Z$ couplings to the internal fermion, while the axial vector
insertion arises from
${\cal O}_{7a}$.
Since the EDM operator is linear in the photon
momentum $q_\mu$, we follow Ref. \cite{Con92} and retain only the terms
linear in $q_\mu$ arising
from this sub-graph.
Denoting its amplitude $T^{\mu\lambda\alpha}$, we choose the loop
momentum routing to satisfy
$q^\mu T_{\mu\lambda\alpha}=0=k^\lambda T_{\mu\lambda\alpha}$, where
$q_\mu$ and $k_\lambda$ are the photon and $Z$-boson momenta, respectively.
The result is the usual anomalous term in
$(q+k)^\alpha T_{\mu\lambda\alpha}$. To linear order in $q$, there exist
three structures
which satisfy these vector current conservation conditions:
\begin{eqnarray}
A^{\mu\lambda\alpha}&=& k\cdot q k_\rho
\epsilon^{\mu\lambda\rho\alpha}-k^\mu\epsilon^{\sigma\lambda\rho\alpha}q_\sigma
k_\rho\\
B^{\mu\lambda\alpha}&=&k^2 q_\nu
\epsilon^{\lambda\mu\nu\alpha}-k^\lambda\epsilon^{\rho\mu\nu\alpha}
 k_\rho q_\nu\\
C^{\mu\lambda\alpha}&=&k^\alpha\epsilon^{\sigma\mu\lambda\rho} q_\sigma
k_\rho\ \ \ .
\end{eqnarray}
The loop integrals for $T^{\mu\lambda\alpha}$ are nominally linearly
divergent. As a check on the calculation, we regulate the
integrals using two different regulators -- Pauli Villars and DR -- and
obtain identical results
in each case. The result is finite:
\begin{eqnarray}
\label{ABJ}
T^{\mu\lambda\alpha}&=& {1\over
8\pi^2}[A^{\mu\lambda\alpha}+5B^{\mu\lambda\alpha}-3
   C^{\mu\lambda\alpha}]\int_0^1\ dx\ {x^2(1-x)\over m_{f'}^2+x(1-x)k^2}\ \ \ .
\end{eqnarray}
It is straightforward to verify that when the corresponding term linear in
$k$ and quadratic in
$q$, the photon momentum\footnote{Obtained from Eq. (\ref{ABJ}) by
$k\leftrightarrow q$,
$\mu\leftrightarrow\lambda$}, is added to the expression in Eq. (\ref{ABJ})
and the divergence
$(q+k)^\alpha T_{\mu\lambda\alpha}$  computed, one obtains the finite,
textbook result for
$k^2=q^2=0$ \cite{IandZ}.

The closed fermion loop of Fig. \ref{fig:twoloop}b
contains axial vector insertions from
${\cal O}_{7a}$
and from the coupling of the $Z$-boson to the internal
fermion. The external
fermion line contains the vector current $Z$-fermion coupling. This
sub-graph diverges
quadratically and must be renormalized by the appropriate $\overline{MS}$
subtraction before
the remaining loop integration is performed. In all cases, the amplitudes
are infrared-finite.
Consequently, we follow Ref. \cite{Con92} and neglect the fermion mass
dependence entering the
loops.

At leading-log order, the results are
\begin{equation}
\label{eq:oneloop}
C_5^f\sim e C_7^{\gamma Z} \left({\mz\over\lamtv}\right)^2\left({1\over s_W
c_W}\right)
   \gaf\left({1\over 16\pi^2}\right) \ln{\mzs\over\mu^2} \ \ \
\end{equation}
for the one-loop contribution in Fig. \ref{fig:oneloop_2quark},
which contains ${\cal O}_{7c}$
\cite{Cor00}. The loops in Fig. \ref{fig:twoloop},
which contain the four-fermion operator ${\cal O}_{7a}$, yield
\begin{equation}
\label{eq:twoloop}
C_5^f\sim - e C_7^{ff'} \left({\mz\over\lamtv}\right)^2
Q_{f'}g_\sst{V}^{f'}\gaf \left({G_\sst{F}
   \mzs\over\sqrt{2}}\right)\left({1\over 8\pi^2}\right)^2
\ln{\mzs\over\mu^2}\ \ \
\end{equation}
for the amplitude of Fig. \ref{fig:twoloop}a and
\begin{equation}
\label{eq:compton}
C_5^f\sim  - e C_7^{ff'} \left({5\over
12}\right)\left({\mz\over\lamtv}\right)^2 Q_f g_\sst{V}^f
g_\sst{A}^{f'} \left({G_\sst{F}\mzs\over\sqrt{2}}\right)\left({1\over
8\pi^2}\right)^2
\left(\ln{\mzs\over\mu^2}\right)^2\ \ \
\end{equation}
for the amplitude of Fig. \ref{fig:twoloop}b.  Here, $s_W$ and $c_W$ denote
the sine
and cosine of the Weinberg angle, respectively, $g_\sst{V}^{f}$ and $\gaf$
denote the
vector and axial vector
couplings of the $Z$-boson to fermion $f$, and $Q_f$ is the corresponding
fermion electromagnetic
(EM) charge\footnote{We have not considered loop effects involving $\osgg$.
Since they contribute to the
EDM at two-loop order, however, we expect them to be no more important than
the two-loop
contributions involving $\osffp$.}

As expected from general considerations, the results in Eqs.
(\ref{eq:oneloop}-\ref{eq:twoloop})
display the $(p/\lamtv)^2$ suppression relative to the tree-level relation
in Eq.
(\ref{eq:edmtree}), where in this case $p=\mz$. Thus, if the $C_7$ have
natural size
\begin{eqnarray}
\label{eq:naturala}
C_{7a}&=& 4\pi\kappa^2 \\
\label{eq:naturalb}
C_{7c}&=& 4\pi\kappa^2 eg = {4\pi\alpha\over s_W}C_{7a}
\end{eqnarray}
and if $\lamtv$ is determined from the experimental limit on $\de$ via
Eq. (\ref{eq:edmtree}), then the
loop corrections to the tree-level EDM will be suppressed by more than
25 orders of magnitude for new strong  interactions ($\kappa^2\sim 1$).
In this case, tree-level dominance of the EDM implies that EDM
limits do not provide direct bounds on the $d=7$ operators appearing in Eq.
(\ref{eq:lnew}).  The situation here is analogous to that of the baryon
magnetic moments.  Since the latter are generally dominated by the tree-level
term, they cannot be used to determine the meson-baryon
couplings which enter the one-loop, sub-leading contributions. Instead,
the meson-baryon
interaction must be determined directly from, {\em e.g.}, $\pi N$
scattering and the results used as input into the chiral loop corrections.

The experimental EDM limits could be used to constrain the $d=7$ TVPC
operators directly if -- as in Scenario (A) --
$C_5^f$ vanishes at tree-level in the EFT. Such a situation could arise
under scenarios,
such as left-right symmetric gauge theories, in which parity symmetry is
restored at some
scale well below $\lamtv$. In this case, there would exist no
short-distance PV interactions
to conspire with the new TVPC physics in generating a tree-level $C_5^f$.
The leading contribution
to an elementary fermion EDM would then be given by the results in Eqs.
(\ref{eq:oneloop}-\ref{eq:compton}), with $\mz$ replaced by, {\em e.g.},
the scale of
parity-restoration (such as the mass of a right-handed gauge boson), and
with the appropriate
combinations of couplings. Since the low-energy scale ({\em e.g.}, $\mz$)
enters quadratically,
a conservative upper bound on the $C_7/\lamtvc$ for this scenario can be
obtained by using the
Standard Model results given above. The most stringent constraint results
from the one-loop
amplitude in Eq. (\ref{eq:oneloop}) applied to the electron EDM. Using the
parameterization of
Eq. (\ref{eq:naturalb}), we obtain $\lamtv\simge 260\ \kappa^{2/3}$ TeV.
The corresponding
two-loop limits are about a factor of five $\sim (16\pi^2)^{1/3}$ weaker.
The one-loop constraints
from the neutron EDM are also slightly weaker, given the somewhat less
stringent experimental
limits on $d_n$ \cite{neutronedm}.

The results of the foregoing analysis have important implications for
low-energy direct TVPC
searches in light quark systems. These implications are most transparent
under Scenario A. In this case, EDM limits constrain the ratio
$C_7/\lamtvc$ via the
one- and two-loop results of Eqs. (\ref{eq:oneloop}-\ref{eq:compton}). As
noted in Ref. \cite{Eng96}, one expects the ratio $\alpha_T$ to scale as
\begin{equation}
\label{eq:alphaT}
\alpha_T\sim C_7\left({p/\lamtv}\right)^3\ \ \ ,
\end{equation}
where $p$ is a momentum characteristic of low-energy hadronic interactions.
Taking $p = 1$
GeV$/c$ and using the one-loop electron EDM results, one obtains a limit of
$\alpha_T \simle 10^{-15}$. By comparison, the present direct TVPC search
limits are considerably weaker:
$\alpha_T\simle 10^{-3}$. In short, under the parity-restoration scenario (A),
the EDM results provide
the most stringent bounds by many orders of magnitude.

For Scenario (B) in which $C_5^f$ does not vanish at tree-level ({\em e.g.},
PV persists at
short distances), the situation is more subtle. In this case the EDM
results do not
provide direct constraints on the $d=7$ operators. Nevertheless, one might
argue from the
EDM limits that the low-energy effects of $d=7$ operators should be
considerably smaller than
the present sensitivity of direct TVPC searches. Comparing Eqs.
(\ref{eq:oneloop}-\ref{eq:compton}) to Eq. (\ref{eq:alphaT}), we infer that
the low-energy
TVPC effects of the $d=7$ operators should be suppressed relative to the
corresponding
contributions to an EDM by
\begin{equation}
\left({p\over \mz}\right
)^2 \ \times \ \left(1/{\hbox{loop factors}}\right)\ \ \ .
\end{equation}
If, in addition, the $d=7$ loop contributions are already suppressed
relative to the
tree-level EDM by many orders of magnitude, one would conclude the that
corresponding effects in
low-energy ($p\lsim 1$ GeV$/c$) TVPC processes would be even smaller --
certainly well below the present direct search sensitivity.

Nevertheless, this line of reasoning is not airtight. If $\lamtv\sim\mz$, it is
conceivable that the tree-level and loop contributions to the EDM can be
comparable in magnitude
and that, due to possible cancellations, the magnitude of either term can
be considerably larger
than the EDM limit itself. An analogous situation arises, for example, in the
chiral expansion of the
isoscalar nucleon magnetic moment. In this case, the leading corrections to
the tree-level
relation in Eq. (\ref{eq:mmoments}) arise from kaon loops. Since $m_K$ and
$\Lambda_\chi$ do not
differ appreciably, the loop corrections are consirably larger than the
isoscalar magnetic moment.
A similarly large short-distance (tree-level) contribution is needed to
cancel the loop effect
and obtain the small isoscalar magnetic moment. In this case, the use of
the isoscalar magnetic
moment, together with power counting and na\"\i ve dimensional analysis, to
infer either the size
of $\Lambda_\chi$ via the tree-level relation Eq. (\ref{eq:mmoments}) or the
size of the one-loop
effects would lead to erroneous conclusions. An independent determination
of the strength of the
loop contribution ({\em e.g.}, of the kaon-baryon interaction) is needed.
Should a similar
situation obtain for the EDM, then direct TVPC searches would still be
needed to ascertain the
scale of the $d=7$ contributions.

Before concluding our discussion of the elementary fermion EDM, we
emphasize the differences
between our analysis and that of Ref. \cite{Con92}. In that work, a
calculation of the amplitudes in Fig. 3 was used
to try and estimate the size of the short-distance contributions. This
estimate was implemented by
regulating the two-loop integrals corresponding to Fig. \ref{fig:twoloop}
with a form
factor of the type
\begin{equation}
F_0(p^2)=(p^2/\lamtvs-1)^{-1}\ \ \ .
\end{equation}
The use of this regulator causes the loop integrals
to be dominated by
contributions from intermediate states having $p\sim\lamtv$, thereby
blurring the separation of
scales crucial to the EFT expansion of Eq. (\ref{eq:lnew}). Consequently,
the two-loop results
of Ref. \cite{Con92} scale, incorrectly, as $1/\lamtv$ rather than as
$1/\lamtvc$. The
corresponding implications for low-energy direct TVPC observables are,
therefore, erroneous.

More generally, as noted in Ref. \cite{MRM99a}, the use of a cut-off
regulator destroys the
power-counting which justifies truncation of the EDM analysis at $d=7$.
This loss of a systematic
expansion can be seen by considering the tower of operators
\begin{equation}
\label{eq:tower}
{\cal O}_{7+2n}^{ff'} =  {\bar\psi}_f {\buildrel \leftrightarrow
\over D_\mu}
\gamma_5 \psi_f (\partial^2)^n
	   {\bar\psi}_{f'}\gamma^\mu\gamma_5 \psi_{f'}\ \ \ ,
\end{equation}
where $n=0,1,\ldots$. The two-loop calculation of Ref. \cite{Con92} may be
repeated by
replacing the insertion of $\osffp$ by each of the operators in Eq.
(\ref{eq:tower}). To
regulate the divergences, one may, as the calculation of Ref. \cite{Con92},
regulate the integrals
with a form factor
\begin{equation}
F_n(p^2)=(p^2/\lamtvs-1)^{-(1+n)}\ \ \ .
\end{equation}
The corresponding loop
integrals will be the
same as those in Ref. \cite{Con92} but with additional factors of
\begin{eqnarray}
\label{eq:binomial}
&&\left(p^2/\lamtvs\right)^n\left(p^2/\lamtvs-1\right)^{-n}\\
&=&\left(p^2/\lamtvs-1\right)^{-n}
\left[(p^2/\lamtvs-1)^n+ n(p^2/\lamtvs)^{n-1}-\cdots\right]\\
\label{eq:binomialc}
&=&1+n(p^2/\lamtvs)^{n-1}(p^2/\lamtvs-1)^{-n} - \cdots\ \ \
\end{eqnarray}
appearing in the integrand.
The first term ($=1$) on the RHS of Eq. (\ref{eq:binomialc}) will yield the
same leading-log
contribution as obtained in the calculation of Ref. \cite{Con92}. The
remaining terms will
generate sub-leading contributions, finite as $\lamtv\to\infty$. Thus, at
leading-log order,
each operator in the tower will generate the {\em same} contribution, apart
from the operator
coefficient $C^{ff'}_{7+2n}$, so that the EDM will be proportional to
\begin{equation}
\label{eq:edmseries}
\sum_{n=0}^{\infty} C^{ff'}_{7+2n} \ \ \ .
\end{equation}
In this case, there exists no reason to isolate the effects of the $d=7$
operators from those of
any other operator in the tower. All contribute with equal weight. It would
be erroneous,
therefore, to truncate the series at $d=7$, as was done in Ref.
\cite{Con92}, and to argue that
the EDM limits constrain the magnitude of only one term in this infinite
series.

As this example illustrates, the preservation of the scale separation is
crucial to the power
counting arguments which justify truncation of the expansion of the EDM at
a given order.
When DR and $\overline{\hbox{MS}}$ subtraction is used, for example, the
contributions of each
operator in the tower (\ref{eq:tower}) will be supressed by successive
powers of $(\mz/\lamtv)^2$.
To the extent that $\mz/\lamtv < 1$, truncation at $d=7$ makes for a
reasonable approximation.
In the remainder of this study, we therefore to work with DR and
$\overline{\hbox{MS}}$
subtraction in treating loop effects.

\bigskip

\section{Four-quark TVPV Operators}

The previous discussion considered the EDM of an elementary fermion.
For a composite system such as a neutron, for example, one
must also consider
many-body contributions involving more than one quark degree of freedom. A
generic contribution
of this type is shown in Fig. \ref{fig:hadronic}. The operators which
describe these
many-quark effects include
the TVPV $d=6, 7$ operators listed in Eqs. (\ref{eq:d6tvpv},
\ref{eq:d7tvpv}). As in the case
of the single fermion EDM ${\cal O}_5^{f\gamma}$, the $d=6,7$ operators may
exist at tree-level in
the EFT if parity is violated at sufficiently high scales (Scenario B).
Similarly,
these operators may be
renormalized by PV radiative corrections to the $d=7$ TVPC operators. As
with ${\cal O}_5^{f\gamma}$, the ${\cal O}_{6,7}^\sst{TVPV}$ will be
dominated by these loop effects if
parity symmetry is restored for $\mu < \lamtv$ (Scenario A). In what
follows, we compute
the relevant loop effects.

The leading corrections to the ${\cal O}_{6,7}^\sst{TVPV}$ are generated by
the set of graphs illustrated in Fig. \ref{fig:oneloop_4quark},
where the operator inserted is $\osffp$. The diagrams of Fig.
\ref{fig:oneloop_4quark}a
were considered previously in
the study of Ref. \cite{Khr91}. The diagrams for Fig.
\ref{fig:oneloop_4quark}b,
which are required by electromagnetic
(EM) gauge invariance, were not included in that analysis. The inclusion of
these graphs is
needed in order to obtain the pieces of the $d=7$ TVPV operators containing
the photon field. As
we discuss in Section 6, the contributions from these $\gamma$-insertion
diagrams to the neutron
EDM are numerically as large as the contributions arising from the graphs
of Fig. \ref{fig:oneloop_4quark}a. The reason
for this situation is relatively straightforward. The
diagrams in Fig. \ref{fig:oneloop_4quark}a yield
the pieces of the $d=7$ operators containing a derivative, as well as
contributions to the
$d=6$ operators proportional to one power of quark mass\footnote{Since the
$d=6$ operators do not preserve chirality, they cannot be induced
dynamically by massless quarks.}. When the derivative operator acts on quarks
inside the hadron, the
result is of order $\simle\Lambda_\sst{QCD}$. The resulting
contribution to the neutron
EDM, as in Fig. \ref{fig:hadronic}a for example, will therefore go as
$\Lambda_\sst{QCD}/\Delta M$, where
$\Delta M$ is the mass difference between the neutron and one of its
excited states ({\em e.g.},
an unbound $\pi^{-} p$ pair). The $\gamma$-insertion diagrams of Fig.
\ref{fig:hadronic}b do not produce
such derivative operators, and their corresponding neutron EDM
contributions contain no
$\Lambda_\sst{QCD}/\Delta M$ factors. To the extent that
$\Lambda_\sst{QCD}/\Delta M$ is
of order one, the magnitude of the two sets of contributions should be
comparable.

The amplitudes, ${\cal M }_{\ref{fig:oneloop_4quark}}$, of Fig.
\ref{fig:oneloop_4quark}
are logarithmically divergent. As before, we regulate the
loops using DR and define the finite results using $\overline{\hbox{MS}}$
subtraction. At
leading-log order, the sum of all twenty six diagrams yields the following
linear combination of the ${\cal O}_{6,7}^{TVPV}$ :
\begin{eqnarray}
\label{Four_Quark_Symbolyc}
 {\cal M }_{\ref{fig:oneloop_4quark}}
 = {C^{ff^\prime}_{7a} \over \Lambda^3_{TVPC}} {\alpha \over 32 \pi s^2_W
c^2_W}
\log \biggl( {\mu^2 \over M_Z^2}\biggr)
\Bigl\{
6 m_{f^\prime}g^{f}_V g^{f^\prime}_A {\cal O}_{6a}^{f f^\prime} \nonumber \\
+ g^{f}_V {\left ( g^{f}_A+g^{f^\prime}_A \right )\over 2}
\Bigl[
3{\cal O}_{7e}^{ff^\prime} + {\cal O}_{7g}^{f^\prime f}
\Bigr] \nonumber \\
+  m_{f^\prime}g^{f^\prime}_A g^{f}_V {\cal O}_{6c}^{f^\prime f}
+ \left( g^{f^\prime}_A g^{f}_V - 2 g^{f^\prime}_A g^{f^\prime}_V -
2g^{f}_A g^{f^\prime}_V \right){\cal O}_{7f}^{f^\prime f}
\nonumber \\
-{3 \over 2} g^{f}_A g^{f^\prime}_V {\cal O}_{7h}^{f^\prime f}
+{1 \over 2} g^{f}_A g^{f^\prime}_V {\cal O}_{7j}^{f^\prime f}
\Bigr\}\ \ \ .
\end{eqnarray}
The expression in Eq. (\ref{Four_Quark_Symbolyc})
is obtained by keeping the external fermion lines off shell. Doing so allows us
to identify uniquely the contributions of the $d=7$ TVPV derivative
operators and verify the
gauge invariance of the overall result.

In order to compare our result with the calculation of Ref. \cite{Khr91},
we use the
equations of motion and let the quarks go on shell and convert to momentum
space.
We obtain
\begin{eqnarray}
\label{On_Shell_Four_Qark_I}
 {\cal M }_{\ref{fig:oneloop_4quark}} = {C_{7a}^{ff^\prime}
 \over \Lambda_{TVPC}^3} {\alpha \over 32 \pi s^2_W  c^2_W }
\log \biggl( {\mu^2 \over M_Z^2}\biggr)
\Bigl[
6im_{f^\prime} g^{f}_V g^{f^\prime}_A
{\bar U}_{f} \gamma_5{U}_{f} {\bar U}_{f^\prime} {U}_{f^\prime}
\nonumber \\
-2im_{f^\prime} g^{f}_V \left ( g^{f}_A + g^{f^\prime}_A \right )
{\bar U}_{f}{U}_{f}
{\bar U}_{f^\prime} \gamma_5{U}_{f^\prime} \nonumber \\
+im_{f^\prime} g^{f^\prime}_A g^{f}_V {\bar U}_{f} \gamma_5
\sigma^{\mu\nu}{U}_{f}
{\bar U}_{f^\prime} \sigma_{\mu\nu} {U}_{f^\prime} \nonumber \\
+i \left ( 2g^{f^\prime}_A g^{f^\prime}_V-g^{f^\prime}_A
g^{f}_V + {7 \over 2} g^{f}_Ag^{f^\prime}_V \right)
{\bar U}_{f} \gamma_5 {\left ( p'_{f} + p_{f} \right )}^\mu {U}_{f}
{\bar U}_{f^\prime} \gamma_\mu {U}_{f^\prime} \nonumber \\
+{1 \over 2} g^{f}_Ag^{f^\prime}_V {\bar U}_{f} \sigma^{\mu\nu}{U}_{f}
{\bar U}_{f^\prime} {\left ( p'_{f^\prime} + p_{f^\prime} \right)}_\nu
\gamma_\mu \gamma_5 {U}_{f^\prime}
\Bigr] \ \ \ ,
\end{eqnarray}
where $U_f\equiv U(p_f)$ and ${\bar U}_{f}\equiv {\bar U}(p_f')$ are the
spinors for
incoming and outgoing fermions $f$, respectively.

The first three of the terms on the RHS of Eq. (\ref{On_Shell_Four_Qark_I})
are identical to those appearing in
\cite{Khr91}. Our coefficient of the fourth term, however, differs from the
corresponding
expression in \cite{Khr91}, and the fifth term does not appear in that work
at all.
We trace part of the difference on the fourth term to diagrams in Fig. 5a
where the $Z^0$ boson
connects to initial and final quarks of the same species. It is unclear
from the discussion of
Ref. \cite{Khr91} whether those authors included this class of diagrams.
Given that our sum of
the amplitudes for Figs. 5a and 5b satisfies a gauge invariance
self-consistency check,
we are confident in our result.

\bigskip

\section{Hadronic Matrix Elements}

The operators obtained in the previous sections can be used to compute
contributions to the EDM of the neutron and of neutral atoms. Doing so
requires that one
calculate various hadronic matrix elements of two- and four-quark
operators. A first
principles treatment of these matrix elements in QCD goes beyond the scope
of the present
study. Moreover, since we seek only to derive order of magnitude
constraints on new
TVPC interactions and not to obtain definitive numerical results, it
suffices to draw
upon various approximation methods. To that end, we turn to the quark model
\cite{DGH86}, factorization, and chiral symmetry.

Below, we estimate a number of different matrix elements relevant to the
neutron
EDM and couplings $\grhobar$ and ${\bar g}_\pi^{(a)}$:

\begin{enumerate}

\item The contribution to $d_n$ from quark EDMs (Fig. 6)

\item The relationship between $\grhobar$ and $d=7$ TVPC operators (Fig.
7a) and
the relationship between $\grhobar$ and $d_n$ (Fig. 7b).

\item The contribution to $d_n$ from the four-quark/photon TVPV operators
appearing
in Eq. (\ref{Four_Quark_Symbolyc}) (Fig. 4b).

\item The contribution to the ${\bar g}_\pi^{(a)}$ from the purely hadronic
terms
in Eq. (\ref{Four_Quark_Symbolyc}) (Fig.8a) and the relationship between
the ${\bar g}_\pi^{(a)}$
and $d_n$ (Fig. 8b).

\item The tree-level contribution from ${\cal O}_{7c}^{Z\gamma}$ (Fig. 9).

\end{enumerate}

\bigskip
\noindent {\bf A. Two-quark  TVPV Operators}

Relating the EDM of a constituent quark to that of the neutron using the
quark model is
a straightforward exercise. As shown in Ref. \cite{MRM99a}, this
relationship is given by
\begin{equation}
\label{eq:qmrel}
d_n={1\over\lamtv}\left[{4\over 3}C_5^d-{1\over 3} C_5^u\right]\int\ d^3x\
\left(u^2+{1\over 3}
\ell^2\right)\ \ \ ,
\end{equation}
where $u$ and $\ell$ are the upper and lower component quark model radial
wave functions,
respectively. The integral in Eq. (\ref{eq:qmrel}) can be estimated using
the wave function
normalization condition
\begin{equation}
\label{eq:wavenorm}
\int\ d^3x\ \left(u^2+\ell^2\right) = 1
\end{equation}
and expression for the axial vector charge
\begin{equation}
\label{eq:axial}
g_A = {5\over 3}\int\ d^3x\ \left(u^2-{1\over 3}\ell^2\right)\ \ \ ,
\end{equation}
where $g_A\approx 1.26$. From Eqs. (\ref{eq:wavenorm}-\ref{eq:axial}) one
obtains
\begin{equation}
\label{eq:qmintegral}
\int\ d^3x\ \left( u^2+{1\over 3}\ell^2\right) = {1\over 4}\left(2+{6\over
5}g_A\right)
\approx 0.88\ \ \ .
\end{equation}

\bigskip
\noindent {\bf B. Four-Quark TVPC Operators}

Deriving the relationship between the effective hadronic coupling
$\grhobar$ and the four-quark
$d=7$ TVPC operators (Fig. 7a) requires more thought than in the case of
evaluating two-quark matrix elements. For simplicity, we focus on the
four-quark operator
${\cal{O}}_{7a}^{ff^\prime}$. We make a simple estimate using
factorization. Doing so requires
use of the Fierz transformed version of this operator, since the
interaction in Eq.
(\ref{eq:lrhonn}) involves only $\rho^{\pm}$. The Fierzed form of the
operator is
\begin{eqnarray}
{\cal{O}}_{7a}^{ff^\prime} = && - \Bigl(
{3 \over 4} \bar{\psi}_f\overleftarrow{\partial}_\nu \psi_{f^\prime}
\bar{\psi}_{f^\prime} \gamma_\nu \psi_f  +
{3 \over 4} \bar{\psi}_f\psi_{f^\prime}
\bar{\psi}_{f^\prime} \gamma_\nu \overrightarrow{\partial}_\nu \psi_f
\nonumber \\
&&
+{3 \over 4} \bar{\psi}_f \gamma_\nu \overleftarrow{\partial}_\nu
\psi_{f^\prime}
\bar{\psi_{f^\prime}} \psi_f +
{3 \over 4} \bar{\psi}_f\gamma_\nu \psi_{f^\prime}
\bar{\psi}_{f^\prime} \overrightarrow{\partial}_\nu \psi_f \nonumber \\
 &&
+{i \over 4} \bar{\psi}_f\overleftarrow{\partial}_\nu \gamma_\mu
\psi_{f^\prime}
\bar{\psi}_{f^\prime} \sigma_{\mu  \nu} \psi_f
+ {i \over 4} \bar{\psi}_f \gamma_\mu \psi_{f^\prime}
\bar{\psi}_{f^\prime} \sigma_{\mu  \nu} \overrightarrow{\partial}_\nu
\psi_f \nonumber \\
&&
- {i \over 4} \bar{\psi}_f \overleftarrow{\partial}_\nu \sigma_{\mu  \nu}
\psi_{f^\prime}
\bar{\psi}_{f^\prime} \gamma_\mu \psi_f
 - {i \over 4} \bar{\psi}_f \sigma_{\mu  \nu} \psi_{f^\prime}
\bar{\psi_{f^\prime}} \gamma_\mu \overrightarrow{\partial}_\nu \psi_f
\nonumber \\
 &&
+ {3 \over 4} \bar{\psi}_f \overleftarrow{\partial}_\nu \gamma_5
\psi_{f^\prime}
\bar{\psi}_{f^\prime} \gamma_\nu \gamma_5 \psi_f
+ {3 \over 4} \bar{\psi}_f \gamma_5 \psi_{f^\prime}
\bar{\psi_{f}^\prime} \gamma_\nu \gamma_5 \overrightarrow{\partial}_\nu
\psi_f \nonumber \\
&&
- {3 \over 4} \bar{\psi}_f \overleftarrow{\partial}_\nu \gamma_\nu \gamma_5
\psi_{f^\prime}
\bar{\psi}_{f^\prime} \gamma_5 \psi_f
- {3 \over 4} \bar{\psi}_f \gamma_\nu \gamma_5 \psi_{f^\prime}
\bar{\psi}_{f^\prime}  \gamma_\nu \gamma_5 \overrightarrow{\partial}_\nu \psi_f
\nonumber \\
&& + {i \over 4}
\bar{\psi}_f \overleftarrow{\partial}_\nu \gamma_5 \sigma_{\mu  \nu}
\psi_{f^\prime}
\bar{\psi}_{f^\prime} \gamma_5 \gamma_{\mu} \psi_f
+  {i \over 4} \bar{\psi}_f \gamma_5 \sigma_{\mu  \nu}  \psi_{f^\prime}
\bar{\psi}_{f^\prime} \gamma_5 \gamma_{\mu} \overrightarrow{\partial}_\nu
\psi_f  \nonumber \\
&&
+ {i \over 4} \bar{\psi}_f\overleftarrow{\partial}_\nu  \gamma_5
\gamma_{\mu} \psi_{f^\prime}
\bar{\psi}_{f^\prime} \gamma_5 \sigma_{\mu  \nu} \psi_f
+ {i \over 4} \bar{\psi}_f\gamma_5 \gamma_{\mu} \psi_{f^\prime}
\bar{\psi}_{f^\prime} \gamma_5 \sigma_{\mu  \nu}
\overrightarrow{\partial}_\nu \psi_f \Bigr)
\nonumber \\
\label{eq:fierzed1}
\end{eqnarray}

In the factorization approximation, one makes the replacement
\begin{equation}
\bra{N'}\bar{q_i}^1 O_1 {q_j}^2 \bar{q_j}^2 O_2 {q_i}^1\ket{N\rho}\to
\bra{0} \bar{q_i}^1 O_1 {q_j}^2 \ket{\rho}\bra{N'} \bar{q_j}^2 O_2
{q_i}^1\ket{N}\ \ \ ,
\end{equation}
where $N$ and $N'$ denote nucleons, $q_i^a$ is the field for a quark of
flavor $a$ and
color $i$, and $\bar{q_i}^1 O_1 {q_j}^2 \bar{q_j}^2 O_2 {q_i}^1$ denotes
any of the products
of quark bilinears appearing in Eq. (\ref{eq:fierzed1}). Note that since
hadrons are color
singlets, one has
\begin{eqnarray}
\bra{0} \bar{q_i}^1 O_1 {q_j}^2 \ket{\rho}&=&{1\over 3}\delta_{ij}
  \bra{0} \bar{q_k}^1 O_1 {q_k}^2 \ket{\rho}\\
\bra{N'} \bar{q_j}^2 O_2 {q_i}^1\ket{N}&=&{1\over 3}\delta_{ij}\bra{N'}
 \bar{q_m}^2 O_2 {q_m}^1\ket{N}\ \ \ ,
\end{eqnarray}
where repeated indices are summed over. Hence, each factorization
contribution contains
a factor of
\begin{equation}
{1\over 3}\delta_{ij}\times {1\over 3}\delta_{ji} = {1\over 3} \ \ \ .
\end{equation}

Since any pieces of the interaction above which involve the $\gamma_5$ will
not give rise to a $\rho$ meson-vacuum matrix element, we are concerned
with only
the first eight terms of the Fierzed interaction. We may also reduce the number
of terms to be evaluated using the equations of motion. Of the resulting
operators,
we keep only those containing no powers of the quark mass, since the latter
generate
significant suppression factors. In the case where $f$ represents an up
quark and
$f^\prime$ represents a down quark, the remaining structures are
\begin{equation}
{1 \over 2}  \bar{u} \overleftarrow{\partial}_\nu d \bar{d} \gamma^{\nu} u +
{1 \over 2} \bar{u} \gamma^{\nu}  d \bar{d} \overrightarrow{\partial}_\nu u -
{i \over 4} \bar{u} \overleftarrow{\partial}_\nu \gamma_{\mu} d
\bar{d} \sigma^{\mu \nu} u +
{i \over 4} \bar{u} \sigma^{\mu \nu} d
\bar{d}   \gamma_\mu \overrightarrow{\partial}_\nu u.
\end{equation}
In the factorization approximation, the matrix elements required are
\begin{eqnarray}
 & & \langle 0 | \bar{u} \overleftarrow{\partial}_\mu d | \rho^- \rangle
\langle n | \bar{d} \gamma^\mu u | p \rangle, \nonumber \\
\label{eq:rhome}
& & \langle 0 | \bar{u}  \gamma^\mu d | \rho^- \rangle
\langle n | \bar{d} \overrightarrow{\partial}_\mu  u | p \rangle\\
& & \langle 0 | \bar{u}  \overleftarrow{\partial}_\nu\gamma_\mu d | \rho^-
\rangle
\langle n | \bar{d} \sigma^{\mu\nu}  u | p \rangle, \nonumber \\
& & \langle 0 | \bar{u} \sigma^{\mu\nu} d | \rho^- \rangle
\langle n | \bar{d} \gamma_\mu \overrightarrow{\partial}_\nu | p \rangle,
\nonumber
\end{eqnarray}
plus the corresponding matrix elements for $\rho^+ n\to p$ (note that in
Eq. (\ref{eq:rhome})
the color indices have been suppressed for simplicity). A detailed
evaluation of
these matrix elements appears in Appendix A. The resulting TVPC $\rho NN$
Lagrangian is
\begin{eqnarray}
\label{eq:todd_rho_lagrang}
{\cal{L}} &=& i \sqrt{2} {C_{7a}^{ud} \over \Lambda_{TVPC}^3}
{m_\rho^2 \over f_\rho} {1 \over 6} \bar{N} \Bigl[
\left( {1.05 \over R_\rho} - {1.293 \over R_n} \right) \gamma^\mu \\
&&+ i \left( 0.176 {R_n \over R_\rho} - 0.122 \right) q_\nu \sigma^{\mu
\nu} \Bigr]
(\tau^- \rho^+_\mu \ - \tau^+ \rho_\mu^-) N
\end{eqnarray}

It is customary to write the standard rho-nucleon Lagrangian and the TVPC
rho-nucleon Lagrangian in the respective forms (see, {\em e.g.} Eq.
(\ref{eq:lrhonn}),
\begin{equation}
\label{eq:smrhonuc}
{\cal{L}} = \sqrt{2} f_\rho \bar{N} (\gamma^\mu + i {\kappa_V \over 2 m_n}
\sigma^{\mu \lambda} q_\lambda)(\tau^- \rho^+_\mu + \tau^+ \rho_\mu^-) N
\end{equation}
\begin{equation}
{\cal{L}}_{TVPC} = \sqrt{2} \grhobar f_\rho {\kappa_V \over 2 m_n} \bar{N}
\sigma^{\mu \lambda} q_\lambda(\tau^- \rho^+_\mu - \tau^+ \rho_\mu^-) N
\end{equation}
where $\kappa_V$ is the anamolous isovector magnetic moment,
$m_n$ is the mass of the nucleon and $f_\rho$ is
the $\rho \bar{N} N$ coupling constant.
By redefining the phase of the rho meson, we can eliminate the
Dirac structure $\gamma^\mu$ in Eq. (\ref{eq:todd_rho_lagrang}).
We begin by writing Eqs.
(\ref{eq:todd_rho_lagrang}) and (\ref{eq:smrhonuc}) together as
\begin{eqnarray}
{\cal{L}}_{\rho NN} =
\sqrt{2} f_\rho \bar{N} \left[ \gamma^\mu ( 1 + {iA \over f_\rho})
+ i {\kappa_V \over 2 m_n} (1+ iB {2 m_n \over f_\rho \kappa})
\sigma^{\mu \lambda} q_\lambda \right] \tau^- \rho^+_\mu N
\nonumber \\
+ \sqrt{2} f_\rho \bar{N} \left[ \gamma^\mu ( 1 - {iA \over f_\rho})
+ i {\kappa_V \over 2 m_n} (1 - iB {2 m_n \over f_\rho \kappa})
\sigma^{\mu \lambda} q_\lambda \right] \tau^+ \rho^-_\mu N
\end{eqnarray}
where
\begin{equation}
A = {1 \over 6}{C_{7a}^{ud} \over \Lambda_{TVPC}^3}
{m_\rho^2 \over f_\rho} \Bigl[{1.05 \over R_\rho}-{1.293 \over R_n} \Bigr]
\end{equation}
\begin{equation}
B = {1 \over 6}  {C_{7a}^{ud} \over \Lambda_{TVPC}^3} { m_\rho^2 \over
f_\rho} \left( 0.176 {R_n \over R_\rho} - 0.122 \right)
\end{equation}

We then observe that since $ 1 + i A / f_\rho \approx \exp( i A / f_\rho)$ and
$ 1 + i B 2 m_n /(f_\rho k_V) \approx \exp[i B / (f_\rho \kappa_V)]$,
the phases of the rho mesons can be redefined as,
$\rho^+_\mu \rightarrow {\tilde \rho}_\mu^+ \exp(-iA /f_\rho)$ and
$\rho^-_\mu \rightarrow {\tilde \rho}_\mu^- \exp(iA /f_\rho)$.
This allows us to rewrite the total Lagrangian as
\begin{eqnarray}
{\cal L} \approx \sqrt{2}
f_\rho \bar{N} \left[ \gamma^\mu +  i {\kappa_V \over 2 m_n}
\sigma^{\mu \nu} q_\nu \right] (\tau^- {\tilde \rho}^+_\mu + \tau^+ {\tilde
\rho}_\mu^-) N \nonumber \\
- \sqrt{2} f_\rho \bar{N} \left({B 2 m_n \over f_\rho \kappa_V} -
{A \over f_\rho}  \right) {\kappa_V \over 2 m_n} \sigma^{\mu \nu} q_\nu
(\tau^- {\tilde \rho}^+_\mu - \tau^+ {\tilde \rho}_\mu^-) N
\end{eqnarray}
The quantity $\bar{g}_\rho$ is then
\begin{eqnarray}
\bar{g}_\rho =
{C_{7a}^{ud} \over 3} \left( { m_n m^2_\rho \over \Lambda_{TVPC}^3} \right)
{1 \over f_\rho^2 \kappa_V}
\left[ \left ( {1.05 \over R_\rho} - {1.293 \over R_n} \right )
{\kappa_V \over 2 m_N}
- \left( 0.176 {R_n \over R_\rho} - 0.122 \right) \right]
\end{eqnarray}
where
\begin{equation}
\left ( {1.05 \over R_\rho} - {1.293 \over R_n} \right )
{\kappa_V \over 2 m_N} - \left( 0.176 {R_n \over R_\rho} - 0.122 \right)
\sim -0.023\ \ \ .
\end{equation}

As calculated in Ref. \cite{horing}, $\grhobar$ may contribute to the
neutron EDM
via the loop diagrams of Fig. 7(b) as well as to the atomic EDM via process
like
those shown in Figs. 1(a,b). The $d_n$ calculation of Ref. \cite{horing}
included
the introduction of form factors at the hadronic vertices in order to
render the
loop integrals finite. The result is
\begin{equation}
\label{eq:grhobardn}
{d_n \over e} = {h_{\pi NN} g_{\rho \pi \gamma} f_\rho \bar{g}_\rho \kappa_V
\over 16 \sqrt{2} \pi^2 m_\rho} {\tilde F}(\mpi, \mrho, \mn, \Lambda)
\end{equation}
where  $g_{\rho \pi \gamma} = 0.4$ and the PV pion-nucleon coupling $h_{\pi
NN}$ is
constrained by the PV $\gamma$-decay of $^{18}$F to lie in the range:
$h_{\pi NN} =
(0.73\pm 2.3)g_\pi$, where $g_\pi=3.8\times 10^{-8}$ characterizes the
strength of the
charged current $\Delta S=0$ hadronic interaction. The function ${\tilde
F}$ depends
on the masses appearing in the loop integral as well as the form factor
cut-off
parameter, $\Lambda$.

We note that the use of Eq. (\ref{eq:grhobardn}) and the experimental
limits on $\dn$
to derive bounds on TVPC interactions entails several ambiguities. First,
the value of
$h_{\pi NN}$ measured in nuclei such as $^{18}$F may differ from the value
appropriate to
the single nucleon or few-nucleon systems. Many-body nuclear effects may
renormalize the
long range PV $NN$ interaction in such a way as to shift the value of the
effective PV
$\pi NN$ coupling from the value appropriate for Eq. (\ref{eq:grhobardn}).
This ambiguity
may be resolved by future experiments, such as the measurement of the
${\vec n}+p\to
d+\gamma$ asymmetry planned at LANSCE\cite{Sno00}.

Second, the use of a form factor to render the loop integral finite can
introduce
considerable ambiguity. The form factor chosen in Ref. \cite{horing} was
taken from
the Bonn potential, with $\Lambda=1.4$ GeV. One may just as well have chosen a
cut-off given by the inverse size of the hadron. The variations due to this
spread of
choices can be significant \cite{MRM97}. Moreover, as argued in Section 3,
the use of cut-off regulators can render one's effective theory devoid of any
systematic power counting, leaving it poorly defined. For these reasons, we
will
not use Eq. (\ref{eq:grhobardn}) to derive limits on the $d=7$ TVPC operators.

\bigskip
\noindent{\bf C. Four-Quark TVPV Operators}

We consider the relationship between the $d=6$ and $d=7$ TVPV operators
and the neutron EDM. We specify here to Scenario (A), in which case the TVPV
operators arise entirely from PV radiative corrections to the TVPC operators,
as in Fig. 5. First, we estimate the contribution from the pieces of
the $d=7$ operators containing the photon field. These contributions can be
understood
diagrammatically as shown in Fig 4(b).
Starting from expression in Eq. (\ref{Four_Quark_Symbolyc}), we see that the
TVPV $\gamma$-four-quark interaction can be written as
\begin{eqnarray}
\label{eq:photonop}
{\cal L}_{eff}^{TVPV, EM} = &&
e {\cal J}_\lambda^{eff} A^\lambda \nonumber \\
=  && e{{C_{7a}^{ff'}  \over \Lambda_{TVPC}^3} {\alpha \over 32 \pi s_W^2
c_W^2 }
\log \biggl( {\mu^2 \over M_Z^2} \biggr)}  \Biggl[
 \nonumber \\ &&
 + \left[g^{f^\prime}_V g^{f}_A - g^{f}_V (g^{f^\prime}_A + g^{f}_A)\right]
{\bar\psi}_{f} \sigma_{\lambda \mu} {\psi}_{f}
{\bar\psi}_{f^\prime} \gamma_5 \gamma_\mu{\psi}_{f^\prime} \nonumber \\
&&
+ g^{f^\prime}_V g^{f}_A  3i{\bar\psi}_{f} \gamma_5 {\psi}_{f}
{\bar\psi}_{f^\prime} \gamma_\lambda {\psi}_{f^\prime}
\Biggr]  A^\lambda\ \ \ .
\end{eqnarray}

For simplicity, we consider the case where $f$ is an up quark and
$f^\prime$ denotes a down quark.  As in our estimate of $\grhobar$,
we can find neutron matrix elements using the quark model.
We identify the appropriate quark model expression for the
electric dipole moment,
\begin{equation}
i \langle n \lambda'| \int d^3 x_3 {\cal J}_0^{eff} | n \lambda \rangle_{QM} =
{d_n \over e} \int d^3p \bar{u}(p,\lambda') \sigma_{03} \gamma_5
u(p,\lambda) | \phi(p)|^2\ \ \  .
\end{equation}
Here, $\ket{n\lambda}$ is a neutron in the $S_z=\lambda$ state, the
$\phi(p)$ are used in the wave packet description of the
momentum eigenstates as discussed in Appendix A, and the $u(p,\lambda)$ are
the
neutron spinors. Choosing $\lambda=\lambda'=1/2$ we have
\begin{equation}
\langle n 1/2 | \int d^3 x_3 {\cal J}_0^{eff} | n 1/2 \rangle_{QM} =
{d_n \over e}\ \ \ .
\end{equation}
We evaluate the two matrix elements needed and find
\begin{equation}
\langle n |\bar{u} \gamma_5 u \bar{d} \gamma_0 d | n \rangle
= i {8 \over 9}  {0.8623 \over 4 \pi R_n^2}
\end{equation}
\begin{equation}
\langle n |\bar{u} \sigma^{0 \mu} u \bar{d} \gamma_5 \gamma_\mu d | n
\rangle =
{4 \over 9} {0.8623 \over 4 \pi R_n^2}\ \ \ .
\end{equation}
Using these matrix elements, we obtain
the following expression for the neutron EDM:
\begin{eqnarray}
{d_n \over e}&=& - {C_{7a}^{ud} \over \Lambda_{TVPC}}
\left( {1 \over \Lambda_{TVPC}^2  R_n^2} \right)
{ \alpha \over 32 \pi s_W^2  c_W^2 }
\log \left({\mu^2 \over M_Z^2} \right) \\
&& \ \  \times{ 0.4 \over 4 \pi} \Bigl[ g^d_V (g^u_A + g^d_A)
 + 5.5 g_A^d g_V^u  \Bigr]
\end{eqnarray}
Since $g^u_A + g^d_A \approx 0$  in the Standard Model\footnote{The sum is
exactly zero at
tree-level.}, this contribution to the
neutron electric dipole moment is approximately proportional to
$g_A^d g_V^u$.

The second way TVPV operators contribute to $d_n$ is by mixing states of
opposite parity into
the neutron ground state (Fig. 4a). The lightest state which may contribute
is the $N\pi$ S-wave.
In the chiral limit, its contribution is dominated by the loop diagrams of
Fig. 8(b). The TVPV
$NN\pi$ couplings are just the $\gpibar$, generated from the TVPV quark
operators as
in Fig. 8(a). The $\gpibar$ also contribute to the atomic EDM via the
meson-exchange
interaction of Fig. 1(c).  In what follows, we relate the  $\gpibar$ to the
purely
hadronic parts of the operators in  Eq. (\ref{Four_Quark_Symbolyc}).

Following Ref. \cite{Khr91}, we carry out the calculation in the factorization
approximation while using the on-shell form of the TVPV operators given in Eq.
(\ref{On_Shell_Four_Qark_I}). First, we consider the TVPV $NN\pi^0$ coupling.
The only Dirac structures
which give rise to  pion-vacuum matrix elements are  $\gamma_5$,
and $\gamma_5 \gamma^\mu$.  Therefore the structure in the third line
does not contribute. The last two structures in Eq.
(\ref{On_Shell_Four_Qark_I}) do not
contribute in the factorization
approximation. This conclusion follows from symmetry arguments, $p_\pi^\mu
p_\pi^\nu
\sigma_{\mu \nu} =0$, and the equations of motion
$p_\pi^\mu \bar{N} \gamma_\mu N = 0$. For the remaining structures the PCAC
relation
\begin{equation}
\label{eq:pcac}
{1 \over 2} \langle 0 | \bar{u} \gamma^\mu \gamma_5 u - \bar{d} \gamma^\mu
\gamma_5 d
| \pi^0 \rangle = i f_\pi p^\mu \exp(-ip\cdot x).
\end{equation}
Taking the divergence of Eq. (\ref{eq:pcac}) and using the consequences of
isospin
symmetry
\begin{equation}
\bra{0} {\bar u}\gamma_5 u+ {\bar d}\gamma_5 d\ket{\pi^0} = 0
\end{equation}
yields the following two matrix elements needed in the calculation:
\begin{equation}
\langle 0 | \bar{u} \gamma_5 u  | \pi^0  \rangle = -i {m_\pi^2 f_\pi \over
{m_u+m_d}} \exp(-ip\cdot x)
\end{equation}
\begin{equation}
\langle 0 | \bar{d} \gamma_5 d  | \pi^0  \rangle = i {m_\pi^2 f_\pi \over
{m_u+m_d}}
\exp(-ip\cdot x)\ \ \ .
\end{equation}

In addition, we require the nucleon matrix elements of the light quark
scalar densities. From the quark model and $\pi N$ sigma term, one obtains
\begin{equation}
\langle n | \bar{u}  u  | n  \rangle = \langle p | \bar{d}  d | p  \rangle
\approx 5
\end{equation}
\begin{equation}
\langle n | \bar{d}  d  | n  \rangle = \langle p | \bar{u}  u | p  \rangle
\approx 6
\end{equation}
Using these results, we obtain for the TVPV
neutral pion nucleon couplings $A{\bar n} \pi^0 n$ and $B {\bar p} \pi^0 p$,
the strength of the couplings are:
\begin{equation}
A = {C_{7a}^{ud} \over 3}  {f_\pi m_\pi^2 \over\Lambda_{TVPC}^3}
{ \alpha \over 32 \pi s_W^2 c_W^2 } \log \left({\mu^2 \over M_Z^2}\right)
{m_d \over {m_u+m_d}} 36  \Bigl[ g_V^u g_A^d +
{10 \over 36} g_V^u (g_A^u + g_A^d) \Bigr]
\end{equation}
\begin{equation}
B = {C_{7a}^{ud} \over 3} {f_\pi m_\pi^2 \over\Lambda_{TVPC}^3}
{ \alpha \over 32 \pi s_W^2  c_W^2 } \log\left({\mu^2 \over M_Z^2}\right)
{m_d \over {m_u+m_d}} 30 \Bigl[ g_V^u g_A^d +
{2 \over 5}  g_V^u (g_A^u + g_A^d) \Bigr]
\end{equation}
Here again, since $g_A^u + g_A^d = 0$ at tree level in the Standard Model,
the terms proportional to
$g_V^u g_A^d$  make the largest contribution.

In order to determine the couplings of the charged pion with the nucleon, we
Fierz transform Eq. (\ref{On_Shell_Four_Qark_I}). The result is listed in
Appendix B.
As discussed in Appndix B, we estimate the strength of the $C {\bar p}
\pi^- n$ coupling
using the first two terms in the Fierz transformed expression.  The
remaining terms do not
contribute in the factorization approximation. In this case, we require the
following matrix
elements:
\begin{equation}
\langle 0 | \bar{u} \gamma_5 d | \pi^-  \rangle = \langle \pi^- | \bar{d}
\gamma_5 u | 0\rangle
= i \sqrt{2} {f_\pi m_\pi^2 \over m_u + m_d};
\end{equation}
\begin{equation}
\langle n | \bar{d}  u  | p  \rangle = \langle p | \bar{u}  d | n  \rangle
\approx 1.
\end{equation}
The coefficient can then be written as,
\begin{eqnarray}
C &=& - \sqrt{2} {C_{7a}^{ud} \over 3} \left({f_\pi m_\pi^2 \over
\Lambda_{TVPC}^3} \right)
{ \alpha \over 32 \pi s^2_W  c^2_W}
\log\left({\mu^2 \over M_Z^2}\right) \nonumber \\
&& \times {m_d \over  m_u + m_d}
\left( 4 g_V^u g_A^d  - {1 \over 2} g_V^u g_A^u + {3 \over 4} g_V^d g_A^u
\right)
\end{eqnarray}
The last piece $3 g_V^u g_A^d / 4$ comes from applying equations of motion
to the
derivative terms, as discussed in Appendix B.

Expressed in terms of A, B and C, the PVTV
pion-nucleon couplings from Eq. (\ref{eq:lpinn}) are
\begin{equation}
\gpibarz = (B - A)/6 + \sqrt{2} C / 3
\end{equation}
\begin{equation}
\gpibaro = (A + B)/2
\end{equation}
\begin{equation}
\gpibart = {(B - A)/6 - {C \over 3 \sqrt{2}}}
\end{equation}

In the chiral expansion of the EDM, the dominant contribution to $d_n$ from
the TVPV $\pi NN$
interaction arises from the loops in Fig. 8(b), where the intermediate state
contains a $p\pi^-$. In this case, only the constant $C$ contributes. The loop
calculation, first performed in Ref. \cite{Cre79}, yields
\begin{equation}
{d_n \over e} =  {C \over \sqrt{2}}   { g_{\pi\sst{NN}} \over 4 \pi^2 m_N}
\log {m_N \over m_\pi}\ \ \ .
\end{equation}
Written explicitly in terms of the TVPC scale, $\Lambda_{TVPC}$, the
neutron electric dipole moment takes the form,
\begin{eqnarray}
{d_n \over e} =   \left( C_{7a}^{ud} \over  \Lambda_{TVPC} \right)
\left({m_\pi^2 \over \Lambda_{TVPC}^2} \right) \left({f_\pi \over m_N}\right)
{1 \over 3}
{ \alpha \over 32 \pi s^2_W  c^2_W}
 { g_{\pi\sst{NN}} \over 4 \pi^2} \log {m_\pi \over m_N} \nonumber \\
\log \left ({\mu^2 \over M_Z^2} \right) {m_d \over  m_u + m_d}
\left( 4 g_V^u g_A^d  - {1 \over 2} g_V^u g_A^u + {3 \over 4} g_V^d g_A^u
\right)\ \ \ ,
\end{eqnarray}
where we have kept only the leading, non-analytic loop contribution
proportional to
$\log (m_\pi / m_N)$ and where $g_{\pi NN}$ is the strong $\pi NN$ coupling.
We note that an evaluation of the analytic contributions has
been performed using sidewise dispersion relations in Ref. \cite{Bar69}. In
addition,
loop contributions involving the $n\pi^0$ intermediate state have been
considered in
Refs. \cite{Val90,He93}.

\bigskip
\noindent{\bf D. Tree-level contributions}

There exists one way in which the $d=7$ TVPC operators may contribute to
$d_n$ without
the consideration of loop effects. As shown in Fig. 9, the operator $\oszg$
generates a
tree-level contribution when the $Z^0$ is exchanged and couples to the
second quark's axial
vector current. Na\"\i vely, one might expect this contribution to compete
with the one-loop
quark EDM generated by $\oszg$ in Fig. 2. The calculation of the process in
Fig. 9 is
tedious but straightforward. Using the quark model, we obtain
\begin{eqnarray}
{d_n\over e} & = & -{1\over\lamtv}\left[{C_{7c}^{Z\gamma}\gad\over
2s_W}\right]\left(\mz\over\lamtv
\right)^2\left[{0.34\over M_\sst{Z}^3\mw R^4_n}\right]\\
\nonumber
&\times&\int_0^{2.04}\ x^2dx\left[\j_0(x)^2j_0'(x)j_1(x)
-j_0(x)j_1(x)^2j_1'(x)-j_0(x)j_1(x)^3/x\right]\\
\nonumber
&\times&\sum_{m,m', \alpha,\ldots}\sigma^3_{mm}\bra{n1/2}
u^{\dag}(m',\alpha') u(m',\alpha)
d^{\dag}(m,\beta') d(m,\beta)\ket{n 1/2}\ \ \ ,
\end{eqnarray}
where the sum runs over all spin ($m, m'$) and color $(\alpha, \beta,
\ldots)$ indices.
Evaluating the integrals and quark model contractions yields
\begin{equation}
\label{eq:dntree}
{d_n\over e}  =  {1\over\lamtv}\left[{C_7^{Z\gamma}\gad\over
2s_W}\right]\left(\mz\over\lamtv
\right)^2\left[{0.162\over M_\sst{Z}^3\mw R^4_n}\right]\ \ \ .
\end{equation}

Comparing the result in Eq. (\ref{eq:dntree}) with one-loop amplitude in
Eq. (\ref{eq:oneloop}),
we observe that the tree-level contribution contains the suppression factor
\begin{equation}
{0.162\over M_\sst{Z}^3\mw R^4_n}\approx 8.64\times 10^{-12}\ \ \ .
\end{equation}
The difference results from the contributions of high-momentum ($p\sim\mz$)
intermediate states
in the loops of Fig. 2. The tree-level process, in contrast, is dominated
by states with momenta
$p\simle\Lambda_\sst{QCD}\sim 0.002\mz$. Since the TVPC operator is has
dimension seven, and
since the momentum transferred through
$Z^0$-boson propagator in Fig. 9 is $\simle\Lambda_\sst{QCD}$, one would
expect a suppression factor
of $(\Lambda_\sst{QCD}/\mz)^4\sim 10^{-11}$ for the process of Fig. 9.

\bigskip
\section{Experimental Constraints}

The foregoing analysis allows us to make a quantitative connection between
various observables and effective coupling constants on the one side and
the EFT of TVPC interactions of quarks and gauge bosons on the other. In this
section, we use those relationships to derive bounds on the TVPC interactions.
It is also instructive to interpret these bounds in terms of the mass scale,
$\lamtv$. To that end, we adopt the parameterization of Eqs.
(\ref{eq:naturala},
\ref{eq:naturalb}) and present constraints in terms of $\lamtv$ and $\kappa$.

In what follows, we use experimental $d_e$ and $d_n$ limits directly as well
as the limits on $\grhobar$ and the ${\bar g}_\pi^{(a)}$ derived from several
sources. The corresponding limits on these quantities are given in Tables 1 and
2. It is also useful to convert our expressions for $\grhobar$ and the
${\bar g}_\pi^{(a)}$ into numerical form. In the case of $\grhobar$ we have
\begin{equation}
\label{eq:grhobarnum}
\grhobar\approx -C_{7a}^{ud} \ (\mz/\lamtv)^3\ \times (2\times 10^{-10})\ \ \ .
\end{equation}
Similarly, the combinations of the ${\bar g}_\pi^{(a)}$ relevant to $d_n$ and
$\datom(^{199}\mbox{Hg})$ are, respectively,
\begin{equation}
\label{eq:gpibarnuma}
{\bar g}_\pi^{(2)}-{\bar g}_\pi^{(0)}\approx-C_{7a}\ (\mz/\lamtv)^3\ \times
(6\times 10^{-12})\ \ \
\end{equation}
for $\dn$ and
\begin{equation}
\label{eq:gpibarnumb}
{\bar g}_\pi^{(0)}+{\bar g}_\pi^{(1)}+2{\bar g}_\pi^{(2)}\approx
-C_{7a}^{ud}\ (\mz/\lamtv)^3\ \times (3\times 10^{-11})
\end{equation}
for $\datom(^{199}\mbox{Hg})$.

\medskip
\noindent{\bf Scenario A.} Under this scenario, the first two terms in Eq.
(\ref{eq:edmexpand})
vanish, so that the leading terms in the expansion of the EDM are the
${\cal O}(1/\lamtvc)$
contributions from the TVPC operators. As in the case of the effective
hadronic couplings, it
is useful to express our relationships between the EDM's and the TVPC
interactions in numerical
form.

For the elementary fermion EDM's arising from the one-loop graphs
containing $\oszg$ (Fig. 2), we
require a choice for $\mu$. Since the typical momentum of a quark in the
neutron is $\sim
\Lambda_\sst{QCD}$, we take $\mu=\Lambda_\sst{QCD}$. The appropriate choice
for $\de$ is probably
smaller, on the order of the typical momentum of an electron bound in a
heavy atom. Since the
$\mu$-dependence of $d_f$  is only logarithmic, however, the precise choice
of scale is not
decisive. To be conservative, we also use $\mu=\Lambda_\sst{QCD}$ for
$\de$. The corresponding
results are
\begin{eqnarray}
\label{eq:denum}
{\de\over e}&\sim& C_{7c}^{Z\gamma}\ (\mz/\lamtv)^3\ \times(4\times
10^{-17}\ \mbox{cm})\\
\nonumber
&\sim& C_{7a}^{ff'}\ (4\pi\alpha/s_\sst{W})(\mz/\lamtv)^3\ \times(4\times
10^{-17}\ \mbox{cm})\\
\nonumber
&\approx& C_{7a}^{ff'}\ (\mz/\lamtv)^3\ \times(8\times 10^{-18}\ \mbox{cm})\\
\end{eqnarray}
and
\begin{eqnarray}
\label{eq:dnnuma}
{\dn\over e}&\sim& (\frac{4}{3}\gad-\frac{1}{3}\gau)\times (0.88)\times
{\de\over e}\\
\nonumber
&\approx& C_{7c}^{Z\gamma}\ (\mz/\lamtv)^3\ \times(6\times 10^{-17}\
\mbox{cm})\\
\nonumber
&\sim& C_{7a}^{ff'}\ (\mz/\lamtv)^3\ \times(1\times 10^{-17}\ \mbox{cm})\ \ \ ,
\end{eqnarray}
where we have used the naturalness relation of Eq. (\ref{eq:naturalb}) to
express
$C_{7c}^{Z\gamma}$ in terms of $C_{7a}^{ff'}$
and where the 0.88 factor in Eq. (\ref{eq:dnnuma})
is the value of the bag model integral in Eqs. (\ref{eq:qmrel}) and
(\ref{eq:qmintegral}).
Since the two-loop effects involving $\osffp$ are suppressed relative to
the one-loop
$\oszg$ contributions, we do not give explicit numerical formulae for the
former.

For the contribution to $\dn$ from the TVPV $\gamma$-four quark interaction
of Fig. 4b,
generated by the loops of Fig. 5b, we find
\begin{equation}
\label{eq:dnnumb}
{\dn\over e}\sim C_{7a}^{ud}\ (\mz/\lamtv)^3\ \times (5\times 10^{-25}\
\mbox{cm})\ \ \ ,
\end{equation}
while the process of Fig. 9 yields
\begin{equation}
\label{eq:dnnumc}
{\dn\over e}\sim C_{7a}^{ff'}\ (\mz/\lamtv)^3\ \times(4\times 10^{-28}\
\mbox{cm})\ \ \ ,
\end{equation}
where we have again used Eq. (\ref{eq:naturalb}).

We now apply these expressions to the experimental limits given in Tables 1
and 2. The results
are listed in Table 3. Clearly, the one-loop elementary fermion EDM's
generated by radiative
corrections to $\oszg$ yield the most stringent bounds on $\lamtv$. Whether
one uses $\dn$ or
$\de$, the lower bounds generated by this mechanism are roughly three
orders of magnitude larger
than those obtained from the many-quark effects. The reason for this
difference is clear.
The one-loop graphs are dominated by high-momentum ($p\sim\mz$)
intermediate states, whereas
the many-quark matrix elements are governed by physics at scales
$p\sim\Lambda_\sst{QCD}$. Since
the dimension of the TVPC operators is greater than that of the EDM by two,
the EDM must contain
at least two powers of the relevant mass scale. Hence, the one-loop $\oszg$
effects should
be at least $(\mz/\Lambda_\sst{QCD})^2\sim 2\times 10^{5}$ larger than the
many-quark effects in
$\dn$.

We emphasize the comparison between the one-loop elementary fermion EDM
limits and those obtained
from $\datom(^{199}\mbox{Hg})$. In the latter case, the strongest bounds
are derived from the
combination of the ${\bar g}_\pi^{(a)}$ appearing in Eq.
(\ref{eq:gpibarnumb}). Since these
couplings scale as $1/\lamtvc$, one would require an improvement of nine
orders of magnitude in
the experimental limits on $\datom(^{199}\mbox{Hg})$ in order for the
atomic EDM to compete with
$\de$ and $\dn$\footnote{An improvement of this magnitude, however, would
likely render the atomic
EDM as the most precise probe of $\dn$.}. We also note that were it not for
the one-loop
elementary fermion limits, the diagrams of Fig. 5b -- required by gauge
invariance but omitted in
Ref. \cite{Khr91} -- would yield the most stringent constraints on new TVPC
interactions under
scenario (A). By far the weakest bounds follow from the extraction of
$\grhobar$ from
$\datom(^{199}\mbox{Hg})$.

The limits in Table 3 can be translated into an expectation for $\alpha_T$
under Scenario (A).
Using the general scaling arguments of Ref. \cite{Eng96} which imply
\begin{equation}
\alpha_T\sim C_{7a}^{ff'} \left({p\over\lamtv}\right)^3 =
C_{7a}^{ff'}\left({\mz\over
\lamtv}\right)^3\left({p\over\mz}\right)^3
\end{equation}
and using the naturalness assumption of Eq. (\ref{eq:naturalb}) we have
\begin{equation}
\label{eq:alphatrel}
\alpha_T\sim (7\times 10^{-16}) \times \left({p\over 1\
\mbox{GeV}}\right)^3 \ \ \ .
\end{equation}
For a low energy TVPC process with $p\sim 1$ GeV, the Scenario (A) EDM
limits of Table 3
imply that the size of the effect should be roughly $10^{-15}$. Current
direct search limits
are at the $10^{-3}$ level for $\alpha_T$.

\medskip
\noindent {\bf Scenario B.} As argued above, experimental EDM limits do not
apply to TVPC interactions
when PV persists at short distances. In this case, one must rely on direct
TVPC tests, from which
limits on $\grhobar$ may be extracted. The most precise determinations of
$\grhobar$ from TVPC
observables are obtained from charge symmetry breaking (CSB) terms in $np$
scattering cross sections
\cite{Sim97} and fivefold correlations (FC) in the transmission of
polarized neutrons through spin
aligned Holmium \cite{Huf97}. The limits on $\grhobar$ obtained from these
observables are given in
Table 2. In addition, a proton-deuteron transmission experiment proposed
for COSY would be sensitive
to $\grhobar$ at the $10^{-3}$ level \cite{COSY98}. Further improvements in
the $np$ CSB limits may
also be possible \cite{Van99}.

The bounds on $\lamtv$ from direct TVPC searches are given in Table 4. At
present, the data imply
new TVPC physics could arise at scales as light as a few GeV. Ideally,
future experiments would push
these bounds closer to the weak scale. Such a sensitivity would be
comparable to the present and
anticipated new physics sensitivities of atomic PV and PV electron
scattering experiments
\cite{MRM99b}. Achieving such sensitivity, however, would be a daunting
task, requiring at least
six orders of magnitude improvement in precision beyond the present state
of the art.

\bigskip
\section{Conclusions}

Measurements of EDM's provide one of the most powerful probes of possible
new physics. Thus
far, null results for $\de$, $\dn$, and $\datom$ lead to tight constraints
on a variety of new
physics scenarios. In this paper, we have studied the implications of EDM's
for TVPC new physics.
We have also developed relationships between effective TVPC and TVPV
hadronic couplings, on the
one hand, and TVPC interactions involving quarks and gauge bosons, on the
other. This relationship
has not been systematically delineated in the past. Consequently, some
confusion about the
respective implications of EDM's and the values of hadronic couplings
extracted from direct TVPC
searches has ensued. We believe that our analysis has helped clarify these
implications.

While the possible origin of TVPC interactions in the context of a
renormalizable gauge theory has
yet to be delineated, one may, nevertheless, address the issue using the
framework of EFT. This
framework affords a systematic method for treating nonrenormalizable
interactions and for carrying
out the associated phenomenology. Since the operators we have treated here
have $d>4$, EFT is the
natural framework for performing our analysis. As emphasized above, a key
ingredient in the application
of EFT to this problem is the maintenance of a scale separation. The
short-distance ($p\simge\lamtv$)
TVPC physics about which we are ignorant is parameterized by the {\em a
priori} unknown coefficients
of the nonrenormalizable effective interactions. Long distance ($p<<\lamtv$) physics may be
treated explicitly in the guise of loops and many-body matrix elem
ents involving degrees of freedom
having masses and momenta below $\lamtv$. As we illustrated for the example
of the cut-off regulator
used in the loop calculations of Refs. \cite{Khr91,Con92}, a failure to
adhere to this scale
separation can lead to disastrous results. Indeed, in that example, the
blurring of scale separation
implies (erroneously) that the EDM is proportional to an infinite series of
unknown coefficients of nonrenormalizable
operators of arbitrarily high dimension. Without the truncation scheme
implied by the EFT scale
separation, one is unable to learn anything from experiment about a given
TVPC interaction (or even
finite set of such interactions).

As argued at the outset of our paper, the EFT scale separation implies one
must consider the low-energy
consequences of TVPC new physics under two scenarios: parity-restoration
below $\lamtv$ (Scenario A)
and parity-restoration above $\lamtv$ (Scenario B).
In the case of Scenario A, the EDM is dominated by PV
radiative corrections to  $d=7$ TVPC operators. The most significant impact
arises from the
EDM of elementary fermions, generated by a one-loop graph involving
$\oszg$. Interpreted in terms
of a mass scale, this mechanism, taken with the experimental limits on
$\de$ and $\dn$, imply that
$\lamtv\simge$ 100-250 TeV when the TVPC new physics is \lq\lq strong"
($\kappa\sim 1$). The impact
of many-quark effects -- which contribute both directly to $\dn$ as well as
to both $\dn$ and
$\datom(^{199}\mbox{Hg})$ via the couplings $\grhobar$ and $\gpibar$ -- is
considerably smaller
than that of the one-loop elementary fermion EDM. Indeed, for the
many-quark mechanism to compete
with that of the elementary fermion EDM, one would require an improvement
in the limits on
$\datom(^{199}\mbox{Hg})$ of nine orders of magnitude. The impact of direct
search limits
is even weaker under this scenario. We conclude, then, that $\de$ and $\dn$
provide the most
powerful probes of new TVPC physics when parity symmetry is restored at
short distances, while
the atomic EDM measurements are better suited to constraining other new
physics scenarios.

Under Scenario B, the implications of EDM's are ambiguous at best. The
existence of parity-violation
at short distances implies that $d\leq 7$ TVPV operators contribute to the
EDM along with the
radiatively corrected $d\geq 7$ TVPC operators. Consequently, no single
TVPV observable constrains
the latter unless one invokes additional assumptions. In contrast, the
relationship between TVPC
observables -- such as CSB $np$ scattering cross sections or FC in neutron
transmission -- is not
clouded by the additional unknown constants which enter the EDM. In this
case, however, the limits
obtained from direct TVPC searches are rather weak: $\lamtv\simge$ 1 GeV
(for $\kappa\sim 1$).
As a long term benchmark, one would ideally search for new TVPC physics at
least up to the weak
scale. Achieving this goal would require an improvement in the precision of
direct TVPC searches
by six orders of magnitude. While achieving such improvement would seem
formidable, any intermediate
progress would constitute a welcome step.

\section*{Acknowledgement}
It is a pleasure to thank D. DeMille, R. Springer, and M. Savage, and Wim
van Oers for useful discussions.
This work was supported in part under U.S. Department of Energy contracts
\#DE-AC05-84ER40150 and \#DE-FG06-90ER40561, and \#DE-FG02-00ER41146
and a National Science Foundation Young Investigator Award.

\newpage

\appendix

\section{Quark model matrix elements}
\label{app:sectionA}

%_______________________________________________________________________

Here, we evaluate the matrix elements appearing in Eq. (\ref{eq:rhome}).
We define the following quantities:  $P=p_N+p_P, q=p_\rho=p_N-p_P$, $\epsilon$
is the $\rho$ polarisation.
Since the TVPC $\rho NN$ Lagrangian is linear in the $\rho$ momentum operator,
we retain only those terms in the matrix element products linear in $q$. The
fourth matrix element product in Eq. (\ref{eq:rhome}) is quadratic in $q$
since the nucleon matrix element vanishes to first order in $q$.
Thus, the last producet does not contribute to $\grhobar$.

For the remaining matrix element products in Eq. (\ref{eq:rhome}), we first
require the $\rho$ to vacuum matrix element, parameterized in the
standard way as
\begin{equation}
\label{eq:vc}
 \langle 0 | \bar{u} \gamma_\mu d | \rho^- \rangle
= \sqrt{2} {m_\rho^2 \over f_\rho} \epsilon_\mu e^{i{\vec p}\cdot{\vec x}}
\end{equation}
where $f_\rho^2 / 4 \pi \approx 2.5$.
The rho matrix element with the derivative may be evaluated
using wave packets in the quark model, using the approach of
Donoghue and Johnson \cite{DGH86}.
We begin by assuming the following
structure for the matrix element,
\begin{equation}
\langle 0 | \bar{u} \overleftarrow{\partial}_\mu d | \rho^- \rangle
= (a \epsilon_\mu + b p_\mu) e^{ipx}
\end{equation}
where $p_\mu$ is the momentum of the rho-meson and $\epsilon_\mu$ is
the polarization vector.  We solve for the coefficients, $a$ and $b$.
The quark model matrix element may be written using wave packets as
\begin{equation}
\langle 0 | \bar{u} \overleftarrow{\partial}_\mu d | \rho^- \rangle_{QM} =
\int {d^3 p^\prime \over 2 \omega_{p^\prime}} \phi(p^\prime)
\langle 0 | \bar{u} \overleftarrow{\partial}_\mu d | \rho^- \rangle
\end{equation}

Multiplying each side of the equation by
$\int d^3x \exp(-i{\vec p}\cdot{\vec x}) / (2 \pi)^3$, results
in
\begin{equation}
\int {d^3x  \over (2 \pi)^3} \ \exp(-i{\vec p}\cdot{\vec x})
\langle 0 | \bar{u} \overleftarrow{\partial}_\mu d | \rho^- \rangle_{QM} =
{\phi(p) \over 2 \omega_p} (a \epsilon_\mu + b p_\mu).
\end{equation}
The expression for $\phi(p)$ may be obtained from the vector current
matrix element Eq. (\ref{eq:vc}),
\begin{equation}
\int {d^3 x \over (2 \pi)^3}
\exp(-i{\vec p}\cdot{\vec x}) \langle 0 | \bar{u} \gamma_\mu d | \rho^-
\rangle_{QM}
= \phi(p) {\epsilon_\mu m_\rho^2 \over f_\rho \omega_p \sqrt{2}}
\end{equation}
We use the $\mu = 3$ component of the above equation to find an expression for
$\phi(p)$.  This, taken together with the vector current matrix
element produces,
\begin{eqnarray}
\int d^3 x e^{-i{\vec p}\cdot{\vec x}}\langle 0 | \bar{u}
\overleftarrow{\partial}_\mu d
| \rho^- \rangle_{QM}
&=& \\
\nonumber
\sqrt{2} f_\rho (a \epsilon_\mu + b p_\mu)&& \int d^3 x e^{-i{\vec
p}\cdot{\vec x}}
\langle 0 | \bar{u} \gamma_3 d | \rho^- \rangle_{QM}
\end{eqnarray}

Upon evaluating both quark model matrix elements, we find $b=0$, while,
\begin{equation}
a = {m_\rho^2 \over f_\rho} {2 \over 3} {\int \left[{ du \over dr} l -
2ul/r -u
{dl \over dr} \right] r^2 dr
/ \int (u^2 - l^2/3) r^2 dr}.
\end{equation}
Here, $u$ and $l$ are the upper and lower components of the Quark Model
wave functions,
respectively.
The final
expression is
\begin{equation}
\langle 0 | \bar{u} \overleftarrow{\partial}_\mu  d | \rho^- \rangle
= -{1.05 \over R_\rho} \sqrt{2} {m_\rho^2 \over f_\rho} i \epsilon_\mu
e^{-i{\vec p}\cdot{\vec x}}.
\end{equation}
In this expression, $R_\rho = 3.3-3.5 \, {\rm GeV}^{-1}$ is the radius of
the rho meson in
the bag model.
In the
following we shall also use $R_n = 1 \, {\rm fm}$ for the bag model radius
of the
nucleon.
The nucleon matrix elements also may be evaluated with the quark model.
\begin{equation}
\langle n | \bar{d} \gamma_\mu u | p \rangle
= \bar{u}_n ( \gamma^\mu + 0.168 i R_n \sigma^{\mu \nu} q_\nu) u_p.
\end{equation}
\begin{equation}
\langle n | \bar{d} \overrightarrow{\partial}_\nu u | p \rangle =
0.634 i \omega \bar{u}_n \gamma_\nu u_p
\end{equation}
where $\omega = 2.04 / R_n$ is quark energy inside the bag.

For the third matrix element
product, we obtain using similar methods and Eqs. (\ref{eq:wavenorm},
\ref{eq:axial})
\begin{eqnarray}
& & {1 \over 2} \langle 0 | \bar{u}( \overleftarrow{\partial}_\mu
\gamma_\nu-
\overleftarrow{\partial}_\nu \gamma_\mu )d | \rho^- \rangle =
i{{5+3g_A} \over {15+3g_A}} {{{m_\rho}^2} \over {2^{1/2} f_\rho}}
(\epsilon_\nu q_\mu - \epsilon_\mu q_\nu ) \nonumber \\
& & \langle n | \bar{d} \sigma^{\mu\nu}  u | p \rangle =
{5 \over 12} (1+{g_A \over 5}) \bar{u_N} \sigma^{\mu\nu} u_P
+\cdots
\nonumber\ \ \ ,
\end{eqnarray}
where the $+\cdots$ denote terms higher order in $q$.

\section{Fierz transformed four-quark operators}

In calculating the TVPV $\pi^{\pm}$-nucleon couplings, we require the Fierz
transformed form of
the amplitude in Eq. (\ref{On_Shell_Four_Qark_I}). The transformed amplitude is

\begin{eqnarray}
\label{eq:Fierz_fourquark1}
& & {\cal M}_{5}^\sst{FIERZ} = {C_{7a}^{f f^\prime}
 \over \Lambda_{TVPC}^3} {\alpha \over 32 \pi s_W^2  c_W^2}
\log \biggl( {\mu^2 \over M_Z^2}\biggr)\Bigl\lbrace \nonumber \\
& & i m_{f^\prime}  \Bigl(4 g_A^{f^\prime} g_V^f  -  {1 \over 2} g_V^f
g_A^f
\Bigr)
\bigr( \bar{U}_f \gamma_5 U_{f^\prime} \bar{U}_{f^\prime} U_f +
\bar{U}_f U_{f^\prime} \bar{U}_{f^\prime} \gamma_5 U_f
\bigr) \nonumber \\
& & +  i m_{f^\prime} \Bigl( 2 g_A^{f^\prime} g_V^f  + {1 \over 2}  g_V^f
g_A^f
\Bigr)
\Bigl(\bar{U}_f \gamma_\mu U_{f^\prime} \bar{U}_{f^\prime}
\gamma^\mu \gamma_5 U_f -
\bar{U}_f \gamma_\mu \gamma_5 U_{f^\prime} \bar{U}_{f^\prime}
\gamma^\mu U_f \Bigr)
\nonumber \\
& & - {i \over 4}  m_{f^\prime} g_V^f (g_A^{f} + g_A^{f^\prime})\bar{U}_f
\sigma^{\mu \nu}
U_{f^\prime} \bar{U}_{f^\prime} \gamma_5
\sigma_{\mu \nu} U_f
+ {i \over 4} m_{f^\prime} g_V^f g_A^{f^\prime}
\bar{U}_f \gamma_5 \sigma^{\mu \nu} U_{f^\prime}
\bar{U}_{f^\prime} \sigma_{\mu \nu} U_f \nonumber \\
& & + i (p^\prime_f + p_f)^\mu ( 2 g_A^{f^\prime} g_V^{f^\prime} -
g_V^{f^\prime} g_A^f + {7 \over 2} g_A^f g_V^{f^\prime})  {1 \over 4} (
\bar{U}_f U_{f^\prime} \bar{U}_{f^\prime}
\gamma_\mu \gamma^5 U_f - \bar{U}_f \gamma_\mu \gamma_5
U_{f^\prime} \bar{U}_{f^\prime} U_{f^\prime} ) \nonumber  \\
& & + i (p^\prime_f + p_f)^\mu ( 2 g_A^{f^\prime} g_V^{f^\prime} -
g_V^{f^\prime} g_A^f + {7 \over 2} g_A^f g_V^{f^\prime}) {1 \over 4}
( \bar{U}_f \gamma_5 U_{f^\prime} \bar{U}_{f^\prime} \gamma_\mu
U_f + \bar{U}_f \gamma_\mu U_{f^\prime} \bar{U}_{f^\prime}
\gamma_5 U_f) \nonumber \\
& & - (p^\prime_f + p_f)^\mu ( 2 g_A^{f^\prime} g_V^{f^\prime} -
g_V^{f^\prime}
g_A^f + {7 \over 2} g_A^f g_V^{f^\prime}) {1 \over 4}
(\bar{U}_f  \sigma_{\mu \nu} \gamma_5  U_{f^\prime}
\bar{U}_{f^\prime} \gamma^\nu U_f
- \bar{U}_f \gamma^\nu U_{f^\prime} \bar{U}_{f^\prime} \sigma_{\mu
\nu} \gamma_5 U_f)
\nonumber \\
& & - (p^\prime_f + p_f)^\mu ( 2 g_A^{f^\prime} g_V^{f^\prime} -
g_V^{f^\prime}
g_A^f + {7 \over 2} g_A^f g_V^{f^\prime} ) {1 \over 4}
(\bar{U}_f \gamma^\nu \gamma_5 U_{f^\prime} \bar{U}_{f^\prime}
\sigma_{\mu \nu} U_f + \bar{U}_f  \sigma_{\mu \nu}
 U_{f^\prime} \bar{U}_{f^\prime} \gamma^\nu \gamma_5  U_f)
\nonumber \\
& & + {3 \over 8}   g_A^f g_V^{f^\prime} i (p^\prime_{f^\prime} +
p_{f^\prime})_\mu
( \bar{U}_f \gamma_\mu U_{f^\prime} \bar{U}_{f^\prime} \gamma_5
U_f
+  \bar{U}_f \gamma_5 U_{f^\prime} \bar{U}_{f^\prime} \gamma_\mu
U_f) \nonumber \\
& & - {3 \over 8}   g_A^f g_V^{f^\prime} i ( p^\prime_{f^\prime} +
p_{f^\prime})_\mu
( \bar{U}_f \gamma_\mu \gamma_5 U_{f^\prime} \bar{U}_{f^\prime} U_f
-  \bar{U}_f U_{f^\prime} \bar{U}_{f^\prime} \gamma_\mu \gamma_5
U_f) \nonumber \\
& & - {1 \over 8}  g_A^f g_V^{f^\prime} (p^\prime_{f^\prime} +
p_{f^\prime})_\mu
( \bar{U}_f \sigma_{\nu \mu} U_{f^\prime} \bar{U}_{f^\prime}
\gamma_\nu \gamma_5 U_f
+ \bar{U}_f \gamma_\nu \gamma_5  U_{f^\prime} \bar{U}_{f^\prime}
\sigma_{\nu \mu} U_f ) \nonumber \\
& & + {1 \over 8} g_A^f g_V^{f^\prime} ( p^\prime_{f^\prime} +
p_{f^\prime})_\mu
( \bar{U}_f \gamma_\nu  U_{f^\prime} \bar{U}_{f^\prime} \gamma_5
\sigma_{\nu \mu} U_f
-  \bar{U}_f \gamma_5 \sigma_{\nu \mu} U_{f^\prime}
\bar{U}_{f^\prime}  \gamma_\nu  U_f )
\Bigr\rbrace
\end{eqnarray}

Only the first two terms on the RHS of Eq. (\ref{eq:Fierz_fourquark1})
contribute in the
factorization approximation. The terms on the second line, such as
$\bar{U}_f \gamma_\mu U_{f^\prime} \bar{U}_{f^\prime} \gamma^\mu \gamma_5
U_f$
will not contribute in this approximation.  The
reason is that the pion-to-vacuum matrix element
will be proportional to the pion momentum, $q$, which equals the change of
the momentum of the nucleon. When contracted with the nucleon matrix
element of the vector
current, the result vanishes by current conservation (neglecting the small
$n$-$p$ mass
difference). The terms on the third  line will not contribute for symmetry
reasons.

The remaining terms involve derivatives on the quark fields.  In each case
the derivative
pair is split, so one derivative acts on a quark field in one quark bilinear
and the second derivative acts on the second.  Some of these terms may be
rewritten using
equations of motion for the on shell quarks.  Doing so yields new
contributions to the first
four lines of the RHS of Eq. (\ref{eq:Fierz_fourquark1}), requiring an
adjustment in the coefficients.
The remaining terms  involve derivative
operators, such as
\begin{equation}
\bar{U}_f p_{f\mu}' U_{f^\prime} \bar{U}_{f^\prime}
\gamma^\mu \gamma_5 U_f\ \ \ \mbox{and}\ \ \
\bar{U}^f p_{f\mu}' \gamma^5 U_{f^\prime} \bar{U}_{f^\prime}
\gamma^\mu  U_f.
\end{equation}
Factorization matrix elements of these operators vanish by either symmetry
or current conservation.

Finally, tensor structures such as
$\bar{U}^f p_{f\mu}' \gamma_\nu U_{f^\prime} \bar{U}_{f^\prime}
\sigma^{\mu \nu} \gamma_5 U_f$ cannot produce a pion-vacuum matrix element,
while
those such as
$\bar{U}_f p_{f\mu}'
\gamma_\nu \gamma_5 U_{f^\prime} \bar{U}_{f^\prime}
\sigma^{\mu \nu} U_f$ will vanish from symmetry considerations.

\vfill
\eject

\begin{table}

\caption{Present experimental limits on electric dipole moments.}

\begin{center}
\begin{tabular}{c c c}
Observable  & Experimental limit ($e$-cm) & Ref.\\
\hline
$d_e$ & $\leq 4\times 10^{-27}$ & \cite{Com94} \\
$\dn$ & $\leq 8\times 10^{-26}$ & \cite{neutronedm} \\
$\datom(^{199}\mbox{Hg})$ & $\leq 1.3\times 10^{-27}$ & \cite{Jacobs} \\
\end{tabular}
\end{center}
\label{tab:limits}

\end{table}

\begin{table}

\caption{Experimental limits on TVPC and TVPV hadronic couplings. The value of
$h_{\pi NN}^\sst{DDH}$ is the \lq\lq best value give in Ref. \cite{DDH}.}

\begin{center}
\begin{tabular}{c c c}
Coupling & Experimental limit  & Observable\\
\hline
$\grhobar$ & $\leq 9.3\times 10^{-3}$ & $\datom(^{199}\mbox{Hg})$ \& atomic
PV \\
$\grhobar$ & $\leq 0.53\times 10^{-3}(h_{\pi NN}^\sst{DDH}/h_{\pi NN})$ &
$\dn$ \\
$\grhobar$ & $\leq 5.8\times 10^{-2}$ & FC in neutron transmission\\
$\grhobar$ &   $\leq 6.7\times 10^{-3}$ & CSB in $np$ scattering \\
$|{\bar g}_\pi^{(2)}-{\bar g}_\pi^{(0)}|$  &   $\leq 5.7\times 10^{-12}$ &
$\dn$ \\
$|{\bar g}_\pi^{(0)}+{\bar g}_\pi^{(1)}+2{\bar g}_\pi^{(2)}|$ & $\leq
1.8\times 10^{-11}$ &
				$\datom(^{199}\mbox{Hg})$ \\
\end{tabular}
\end{center}
\label{tab:limits}

\end{table}

\begin{table}

\caption{Limits on new TVPC interactions derived from EDM's under Scenario
(A).}

\begin{center}
\begin{tabular}{c c c}
Observable  &Lower bound on $\lamtv/\kappa^{2/3}$ (TeV) & Mechanism\\
\hline
$\de$ & $ 260$ & single electron $\oszg$ loop \\
$\dn$ & $ 110$ & single quark $\oszg$ loop \\
$\dn$ & $ 0.39$ & Figs. 4b, 5b \\
$\dn$ & $ 0.21$ & $\pi$-loop, Fig. 8b \\
$\dn$ & $ 0.036$ & Fig. 9 \\
$\datom(^{199}\mbox{Hg})$ & $ 0.25$ &  ${\bar g}_\pi^{(a)}$ (Fig. 1c) \\
$\datom(^{199}\mbox{Hg})$ & $ 0.0006$ & $\grhobar$, atomic PV (Fig. 1b) \\
\end{tabular}
\end{center}
\label{tab:limits}

\end{table}

\begin{table}

\caption{Limits on new TVPC interactions derived from direct searches under
Scenario (B).}

\begin{center}
\begin{tabular}{c c c}
Coupling  & Lower bound on $\lamtv/\kappa^{2/3}$ (TeV) & Observable\\
\hline
$\grhobar$ & $ 3\times 10^{-4}$ & CSB in $np$ scattering \\
$\grhobar$ & $ 7\times 10^{-4}$ & FC in neutron transmission\\
\end{tabular}
\end{center}
\label{tab:limits}

\end{table}

\vfill
\eject

\clearpage

\newpage

\begin{figure}

\caption{\label{fig:atomic}
Representative contributions to atomic EDM: (a) TVPV nuclear effect involving
TVPC and PV $\rho NN$ interactions; (b) TVPC nuclear effect plus atomic PV; (c)
long-range TVPV nuclear effect involving TVPV $\pi NN$ coupling. Open
circle denotes
strong meson-nucleon coupling; crossed circle gives TVPC coupling; open square
is TCPV coupling; and crossed square is TVPV coupling.}

\end{figure}

\begin{figure}

\caption{\label{fig:oneloop_2quark}
One loop contribution to elementary fermion EDM from the TVPC operator
${\cal{O}}^{\gamma Z}_7$.
Coupling symbols are as in Fig. 1. }

\end{figure}

\begin{figure}
\caption {\label{fig:twoloop}
Two loop contributions to the elementary fermion EDM
involving the TVPC operator ${\cal{O}}^{f f^\prime}_7$. Symbols are as in
Fig. 1.}

\end{figure}

\begin{figure}
\caption{\label{fig:hadronic}
Contributions from four-quark TVPV operators to $\dn$: (a) second order
contribution
involving a mixture of opposite parity states $\ket{n}$ into neutron; (b)
first-order
contribution arising from $\gamma$-four quark TVPV operators.  Symbols are
as in Fig. 1.}
\end{figure}

\begin{figure}

\caption{\label{fig:oneloop_4quark}
PV weak radiative corrections to $\osffp$, generating $d=7$ TVPV operators.
Symbols are as in Fig. 1.}

\end{figure}

\begin{figure}

\caption{\label{fig:elementaryedm}
Contribution to $\dn$ from individual quark EDM's. Symbols are as in Fig. 1.}
\end{figure}

\begin{figure}

\caption{\label{fig:factorization}
(a) Contributions to $\grhobar$ from $\osffp$; (b) contribution to $\dn$
arising
from TVPC $\rho NN$ and PV $\pi NN$ interactions. Symbols are as in Fig. 1.}

\end{figure}

\begin{figure}
\caption{\label{fig:pionloop}
(a) Contributions to $\gpibar$ arising from TVPV many-quark operators; (b)
leading-oder
contribution (in $\mpi$) to $\dn$ arising from TVPV $\pi NN$ interaction.
Symbols are as in Fig. 1.}

\end{figure}

\begin{figure}

\caption{\label{fig:fourfermionZ}
Tree-level, many-quark contribution to $\dn$ generated by $\oszg$. Symbols
are as in Fig. 1.}

\end{figure}


\begin{thebibliography}{99}

\bibitem{PDG00} The Particle Data Group, Euro. Phys. Journal {\bf C15}, 1
(2000).

\bibitem{Ben99} S.C. Bennett and C.E. Wieman, Phys. Rev. Lett. {\bf 82},
2484 (1999);
   C.S. Wood {\em et al.}, Science {\bf 275}, 1759 (1997).

\bibitem{Der00} A. Derevianko, Phys. Rev. Lett. {\bf 85}, 1618 (2000).

\bibitem{MRM99b} M.J. Ramsey-Musolf, Phys. Rev. {\bf C60}, 015501 (1999).

\bibitem{MRM00a} M.J. Ramsey-Musolf, Phys. Rev. {\bf D62}, 056009 (2000).

\bibitem{Ang98} A. Angelopoulos {\em et al.} (CPLEAR Collaboration), Phys.
Lett.
   {\bf B444}, 43 (1998).

\bibitem{Bla83}E. Blanke {\em et al.}, Phys. Rev. Lett. {\bf 51}, 355 (1983).

\bibitem{Boe95}F. Boehm in {\sl Fundamental Interactions in Nuclei}, W. Haxton
   and E. Haxton, Eds. (World Scientific, Singapore, 1995) p. 67.

\bibitem{Her92}P. Herczeg, J. Kambor, M. Simonius, \lq\lq Parity conserving
time reversal violation in flavor conserving quark-quark interactions", Los
Alamos Theory Division report (unpublished); P. Herczeg, Hyp. Interact.
{\bf 75}
(1992) 127; P. Herczeg in {\sl Fundamental Interactions in Nuclei}, W. Haxton
   and E. Haxton, Eds. (World Scientific, Singapore, 1995) p. 89.

\bibitem{Huf97}P.R. Huffman {\em et al.}, Phys. Rev. {\bf C55}, 2684 (1997).

\bibitem{Koe91}J.E. Koester {\em et al.}, Phys. Lett. {\bf B267}, 23 (1991).

\bibitem{Sim97} M. Simonius, Pys. Rev. Lett. {\bf 78}, 4161 (1997).

\bibitem{COSY98} COSY Proposal, P.D. Eversheim, spokesperson; F. Hinterberger
  (for the TRI Collaboration at COSY), Contribution to the Fifth Annual
	  WEIN Symposium: A Conference on Physics Beyond the Standard Model,
   nucl-ex/9810013.

\bibitem{Van99} W.T.H. van Oers, hep-ph/9912454, to appear in Int. J. Mod.
Phys. {\bf E}.

\bibitem{Lis00}L.J. Lising {\em et al.} (emiT Collaboration), Phys. Rev.
{\bf C62},
  055501 (2000).

\bibitem{Khr91}I.B. Khriplovich, Nucl. Phys. {\bf B352} (1991) 385.

\bibitem{Con92}R.S. Conti and I.B. Khriplovich, Phys. Rev. Lett. {\bf 68}
(1992) 3262.

\bibitem{Eng96}J. Engel, P.H. Frampton, R.P. Springer, Phys. Rev. {\bf D53}
(1996) 5112.

\bibitem{Her95}P. Herczeg, \lq\lq Time reversal violation in nuclear
processes", in {\sl Symmetries and Fundamental Interactions in Nuclei},
W.C. Haxton and
E.M. Henley, eds., World Scientific, Singapore, pp. 89-126.

\bibitem{MRM99a} M.J. Ramsey-Musolf, Phys. Rev. Lett. {\bf 83}, 3997 (1999);
(E) {\em ibid} {\bf 84} 5681 (2000).

\bibitem{MRM00b} M.J. Ramsey-Musolf, hep-ph/0010023, to appear in proceedings
of the Workshop on Fundamental Physics with Pulsed Neutron Beams, Durham, N.C.,
May, 2000.

\bibitem{MRM97} M.J. Ramsey-Musolf and H. Ito, Phys. Rev. {\bf C55}, (3066).

\bibitem{IandZ} C. Itzykson and J.-B. Zuber, Quantum Field Theory,
McGraw-Hill, New York, 1980.

\bibitem{Cor00} Note that the result in Eq. (\ref{eq:oneloop}) is a factor of
six larger than Eq. (6) of Ref. \cite{MRM99a}, which contains an error.

\bibitem{DGH86} J.F. Donoghue, E. Golowich, and B. Holstein, Phys. Rep.
{\bf 131}, 319 (1986).

\bibitem{sak} Sakuri, J. J., Currents and Mesons, University of Chicago
press, Chicago, 1969.

\bibitem{horing} W. C. Haxton, A. Horing, and M. J. Ramsey-Musolf, Phys Rev
{\bf D50} 3422.

\bibitem{Sno00}W.M. Snow {\em et al.}, Nucl. Inst. Meth. {\bf A440}, 729
(2000).

\bibitem{neutronedm} K. F. Smith {\it et al.} Phys. Lett. {\bf B234} (1990)
191,
I. S. Altev {\it et al.} Phys. Lett. {\bf B276} (1992) 242,
P. G. Harris {\it et al.}  Phys. Rev. Lett. {\bf 82} (1999) 904.

\bibitem{Cre79}R.H. Crewther, P. Di Vecchia, G. Veneziano, and E. Witten,
Phys. Lett.
{\bf B88}, 123 (1979); {\em ibid} {\bf 91}, 487 (E) (1980).

\bibitem{Bar69}G. Barton and E.G. White, Phys. Rev. {\bf 184}, 1660 (1969).

\bibitem{Val90}G. Valencia, Phys. Rev. {\bf D41}, 1562 (1990).

\bibitem{He93}X-G. He and B. McKellar, Phys. Rev. {\bf D47}, 4055 (1993).

\bibitem{Com94}E.D. Commins {\em et al.}, Phys. Rev. {\bf A50}, 2960 (1994).

\bibitem{DDH} B. Desplanques, J.F. Donoghue, and B. Holstein, Ann. Phys. (NY)
  {\bf 124}, 449 (1980).
  
\bibitem{Jacobs} J.P. Jacobs {\it et al.}  Phys. Rev. {\bf A52}, 3521 (1995).

\end{thebibliography}
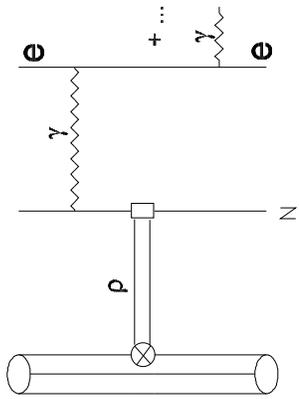
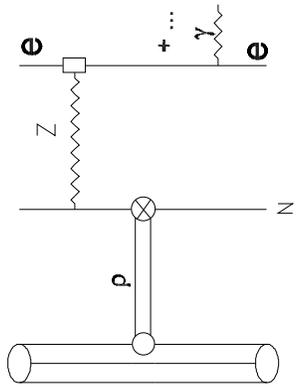
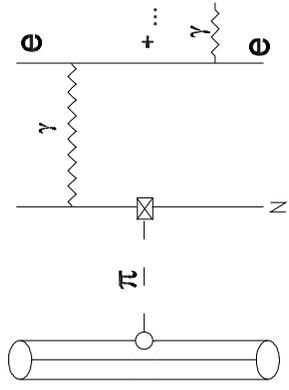
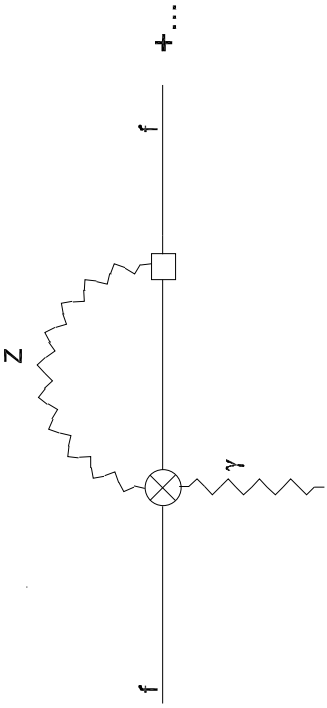
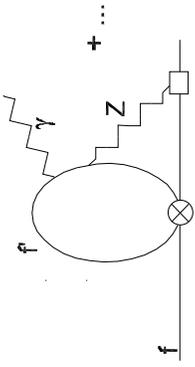
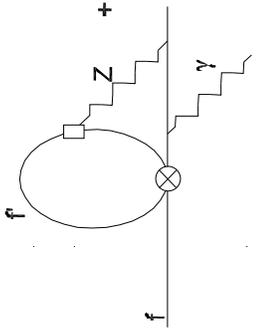
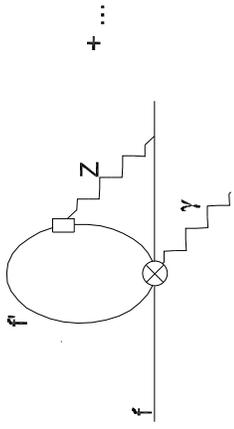
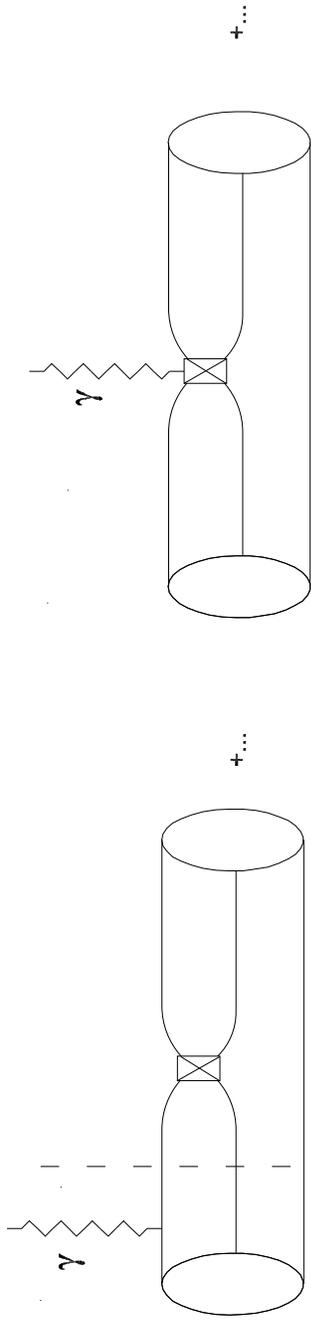
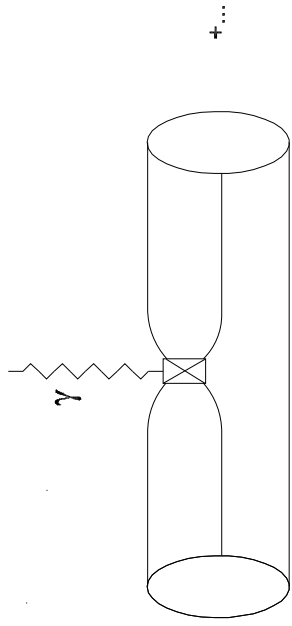
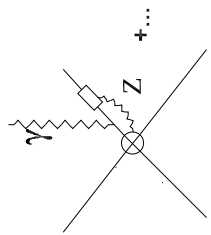
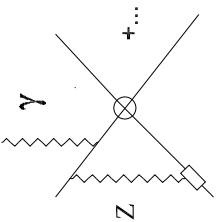
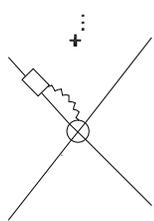
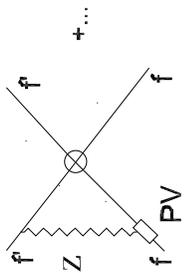
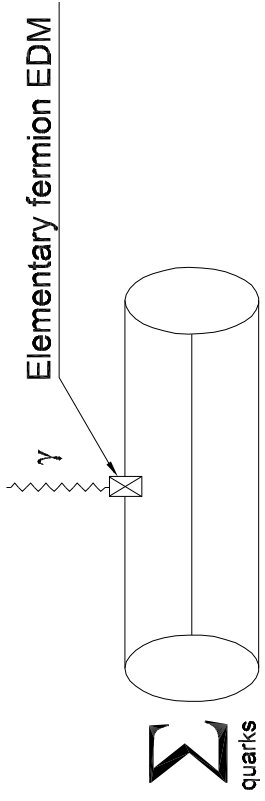
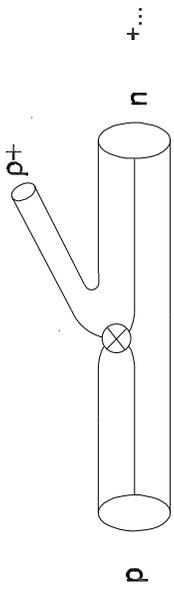
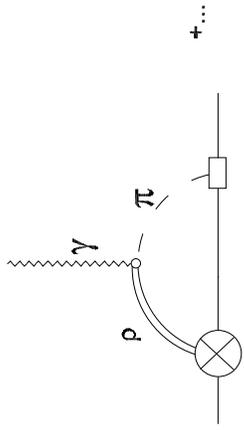
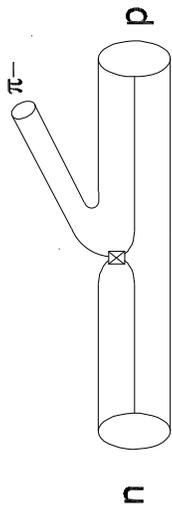
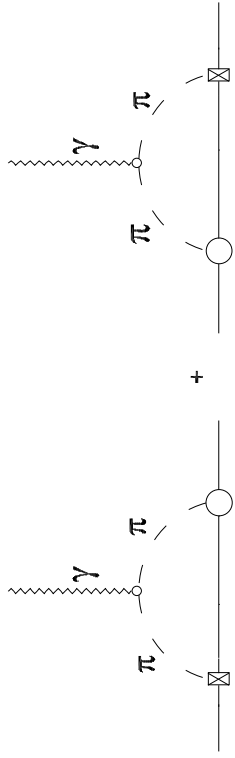
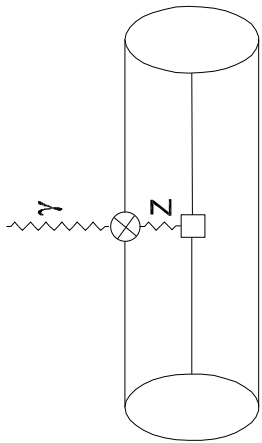
\end{document}